\documentclass[aps,showpacs,twocolumn,amsmath,amssymb,nofootinbib]{revtex4-1}
\usepackage{graphicx,epsfig}
\pdfoutput=1
\usepackage{amsfonts}
\usepackage{mathrsfs}
\usepackage{multirow}
\usepackage{color}
\usepackage{graphicx}  

\def\Journal#1#2#3#4{{#1} {\bf #2}, #3 (#4)}

\def\EPJC{{Eur. Phys. J.} C}
\def\HPA{Helv. Phys. Acta}

\def\PLB{{Phys. Lett.}  B}

\def\PRL{Phys. Rev. Lett.}
\def\PRD{{Phys. Rev.} D}

\def\CQG{{Class. Quant. Grav.}}

\def\GRG{{Gen. Relativ. Gravit.}}

\newcommand{\be}{\begin{equation}}
\newcommand{\ee}{\end{equation}}
\newcommand{\bea}{\begin{eqnarray}}
\newcommand{\eea}{\end{eqnarray}}

\begin{document}

\title{Regular scalar charged clouds around a Reissner-Nordstrom black hole and no-hair theorems}

\date{\today}


\author{Gustavo Garc\'ia}
\email{gustavo.garcia@correo.nucleares.unam.mx} 
 \author{Marcelo Salgado}
\email{marcelo@nucleares.unam.mx} \affiliation{Instituto de Ciencias
Nucleares, Universidad Nacional Aut\'onoma de M\'exico,\\
 A.P. 70-543, CDMX 04510, M\'exico}

\begin{abstract}
  In this work we reanalyze the possibility of finding bound states ({\it scalar clouds}) of a 
  {\it test}, charged and complex-valued scalar field with mass $\mu$ and charge $q$ in the background of a Reissner-Nordstrom black hole (RNBH).
  In order to determine the existence of such scalar clouds we impose suitable
  regularity conditions for the scalar field at the event horizon. We find numerical evidence for the absence
  of such clouds in the subextremal and extremal RNBH when the field is massive but not self-interacting.
  More importantly, we put forward a theorem that
  proves that such clouds cannot exist. On the other hand, when a suitable self-interacting potential is included,
  the theorem no longer applies, providing a heuristic justification behind the existence of charged clouds
  (Q-clouds) that were reported recently.
 \end{abstract}

 \maketitle

\section{Introduction}
\label{Introduction}

In a previous investigation \cite{Garcia}, we presented numerical solutions that represent {\it bound states} of a complex-valued massive scalar field $\Psi$ 
in the test field limit around a subextremal Kerr black hole (BH) of mass $M$ and angular momentum per unit mass $a$. This type of solutions, dubbed {\it clouds}, were found originally by Herdeiro and Radu \cite{Herdeiro2014,Herdeiro2015} in the
test field limit and also when taking into account the backreaction of the field in the spacetime.
More recently in \cite{Garcia2},
we extended the analysis of \cite{Garcia} by considering clouds in extremal Kerr black holes $(a = M)$. For the latter,
it was necessary to consider the extremality condition exactly and not in the limit when $a\rightarrow M$, as in this limit the
boundedness of the radial derivative of the field was not secured. Thus, the extremal case required a separate treatment and
different regularity conditions. These {\it superregular} conditions are different from those considered in the past by
Hod \cite{Hod2012} in that we demanded boundedness on the radial derivatives at the horizon in addition of regularity of the
field itself. The exact solutions found by Hod in the extremal Kerr BH \cite{Hod2012} were further extended by the author to the
extremal Kerr-Newman black hole \cite{Hod2015} by considering a charged and massive field $\Psi$. In this direction
it is worth mentioning the analysis in \cite{Benone2014} where the authors report 
numerical solutions of charged scalar clouds around subextremal Kerr-Newman BH's. Prior to those solutions,
Degollado $\&$ Herdeiro \cite{DegolladoHerdeiro2013} had shown that
it is possible to find scalar clouds around an extremal Reissner-Nordstrom BN (RNBH) only when
the mass $\mu$ of the scalar field $\Psi$ turns to be equal to its electric charge (i.e. $\mu = |q|$),
dubbed {\it double extremal} limit. From that analysis
it seems that the non-trivial solution is possible if the boundedness condition for the radial derivative of the field is dropped.
Otherwise the solution reduces to the trivial one $\Psi=0$ in the domain of outer communication of the RNBH.

In this work we reanalyze the possibility of finding non-trivial charged scalar clouds in the subextremal and extremal 
RNBH when the boundedness of the field and its radial derivatives are imposed at the horizon along the lines proposed in our
previous works \cite{Garcia,Garcia2}. The numerical analysis shows that such clouds do not exist when the
field is massive but not self-interacting. Moreover, we
put forward a (no-hair) theorem based on standard techniques which shows that such clouds cannot exist even when
the boundedness condition on the radial derivatives at the horizon is dropped,
while keeping the scalars formed from the derivatives of the field
regular, casting doubts on the significance of the purported {\it regular} configurations found in
\cite{DegolladoHerdeiro2013}. Finally, we consider a similar scenario but taking into account a self-interacting potential for
the field. In this case, the no-hair theorem no longer applies, which allows us to understand heuristically the
existence of the so-called Q-clouds that were reported lately by several authors~\cite{Hong2020,Herdeiro2020,Brihaye}.

\section{The boson clouds}
\label{sec:clouds}
We assume a RNBH described by the usual metric given in {\it area} coordinates
\footnote{We use units where $G=c=1$.}:
\begin{eqnarray}
\label{RN} 
\nonumber  ds^2 = && -\left(1 - \frac{2M}{r} + \frac{Q^{2}}{r^{2}}\right)dt^{2} + \left(1 - \frac{2M}{r} + \frac{Q^{2}}{r^{2}}\right)^{-1}dr^{2} \\  
& & + \: r^{2}d\theta^{2} + r^{2}\sin^{2}\theta d\varphi^{2}, 
\end{eqnarray}
where $M$ is the {\it mass} and $Q$ is the {\it charge} associated with the RNBH. Under these coordinates,
\begin{equation}
\label{horizons}
r_{\pm} = M \pm \sqrt{M^{2} - Q^{2}} \;,
\end{equation}
provide the location of the external $(r_+)$ and the internal horizon $(r_-)$ of the black hole, where the metric has
coordinate singularities. We will be interested solely
in solving the differential equation for the boson field $\Psi$ in
the domain of outer communication of the RNBH while providing regularity conditions
for $\Psi$ at $r_H \equiv r_+$, in particular, in the extremal case $r_H^{\rm ext}=r_+=r_-=M=|Q|$.

We consider a complex-valued, massive, charged scalar field $\Psi$ which has the following energy-momentum tensor (EMT)
\begin{eqnarray}
\label{Tab}
T_{ab} &=& \frac{1}{2}\Big[\left(D_a\Psi\right)^*\left(D_b\Psi\right) + \left(D_b\Psi\right)^*\left(D_a\Psi\right)\Big] \nonumber \\
&-& g_{ab}\Big[\frac{1}{2}g^{cd}\left(D_c\Psi\right)^*\left(D_d\Psi\right) + U(\Psi^*\Psi)\Big]  \;, 
\end{eqnarray}
where $D_a \equiv \nabla_a - iqA_a$ stands for the covariant derivative associated with the gauge field $A_a$, which in the present case
is given in terms of the electric potential
\begin{equation}
  \label{elecpot}
  A_a= -\Phi (dt)_a= -Q/r (dt)_a \,.
\end{equation}
associated with the RNBH solution; $q$ is the gauge coupling (i.e. electric charge) for the field $\Psi$.
The potential for a free massive field is given by Eq.(\ref{potentialfunction}) provided below, but later
in Sec.~\ref{sec:Qclouds} we analyze a scenario with
self-interaction terms. The EMT (\ref{Tab}) is invariant under the $U(1)$ {\it local} symmetry and
the field $\Psi$ obeys the Klein-Gordon equation 
\begin{equation}
\label{KG}
D^aD_a\Psi=
\left(\nabla^{a} - iqA^{a}\right)\left(\nabla_a - iqA_a\right)\Psi = 2\frac{\partial U\left(\Psi^*\Psi\right)}{\partial \Psi^*}\;,
\end{equation}
where
\begin{equation}
  \label{potentialfunction}
U(\Psi^*\Psi) = \frac{1}{2}\mu^2\Psi^*\Psi.
\end{equation}

We are interested in finding ``bound states" solutions and consider a scalar field $\Psi(t, r, \theta, \varphi)$ 
with temporal and angular dependence of the form, 
\begin{equation}
\label{Psians}
\Psi(t,r,\theta,\varphi)= \phi(r,\theta)e^{i (-\omega t + m \varphi)}\;,
\end{equation}
where $\phi(r,\theta)$ is a real-valued function, and $m$ is a positive integer.
The bound states correspond to a real-valued  frequency $\omega$ equal to the {\it critical} frequency $\omega_{c} \equiv q\Phi_H$ \cite{DegolladoHerdeiro2013}:
\begin{equation}
\label{fluxcond2}
\omega = \omega_c = q\Phi_H\;,
\end{equation}
where $\Phi_H$ is the electric potential at the horizon $r_H$: 
\begin{equation}
\label{OmH}
\Phi_H = \frac{Q}{r_H}\;.
\end{equation}

\section{The subextremal RNBH and regularity conditions}
\label{sec:nonextremal}
In order to solve the Klein-Gordon Eq.~(\ref{KG}) for a free field we assume a mode expansion in the form
\begin{equation}
\label{mode}
\Psi_{nlm}\left(t, r, \theta, \varphi\right) = R_{nlm}\left(r\right)S_{lm}\left(\theta, \varphi\right)e^{-i\omega t},
\end{equation} 
where the angular functions $S_{lm}(\theta, \varphi)$ obey the angular equation
\begin{equation}
 \label{angularE}
\frac{1}{\sin\theta}\frac{d}{d\theta}\left(\sin\theta\frac{dS_{lm}}{d\theta}\right) + \left(K_{l} - \frac{m^2}{\sin^2\theta}\right)S_{lm} = 0 \;,
\end{equation}
where $K_{l}$ are separations constants that relates the radial and angular parts of the Klein-Gordon equation (\ref{KG}).
We observe that Eq.~(\ref{angularE}) corresponds to the Legendre equation, whose solutions are the spherical harmonics $Y_{l}^{m}(\theta, \varphi)$,
and the separation constants $K_{l}$ are provided by
\begin{equation}
\label{constants}
K_{l} = l(l + 1),
\end{equation}
where $l$ is a positive integer. We stress that in this scenario the separation constants do not depend on the {\it magnetic} number $m$,
in contrast with clouds solutions around a Kerr BH~\cite{Herdeiro2014,Herdeiro2015,Garcia}.

Since the separation constants do not depend on the integer $m$ we can change the notation of the radial function $R_{nlm}$ that appears in the Eq.~(\ref{mode}) by the form $R_{nl}$, to describe the radial functions that obey the radial Teukolsky equation \cite{Teukolsky}:
\begin{equation}
\label{radialE}
\Delta\frac{d}{dr}\left(\Delta\frac{dR_{nl}}{dr}\right) + \left[\mathcal{H}^2 - \left(
  K_{l}+ \mu^2r^2 \right)\Delta\right]R_{nl} = 0 \;, 
\end{equation}

where
\begin{equation}
 \Delta = r^{2} - 2Mr + Q^{2} \;,
\end{equation} 
and
\begin{equation}
\label{H}
\mathcal{H} \equiv \omega r^{2} - qQr = \frac{qQr^{2}}{r_H} - qQr = qQr\left(\frac{r}{r_H} - 1\right)\;,
\end{equation}
where we used Eqs.~(\ref{fluxcond2}) and (\ref{OmH}). Notice that $\mathcal{H}$ vanishes at $r = r_H$.

Like in quantum mechanics,
the integer parameters $(n, l, m)$ used to label the scalar-field configurations correspond respectively to
the number of nodes, $n\geq 0$,  for the radial function $R_{nl}$, the {\it angular momentum} $l\geq 0$,
and finally  the ``magnetic" number $m$ satisfies $|m| \leq l$. Given that the background spacetime is
spherically symmetric, intuitively one would not expect the existence of cloud configurations with an angular
dependence, for instance, a dependence on $l$. Nevertheless, we keep this dependence explicitly without assuming
the value $l=0$ in the radial equation for $R_{nl}$.

In order to find configurations that represent {\it bound states} we assume that asymptotically the scalar field
vanishes sufficiently fast. From (\ref{radialE}) one obtains that for $r_H\ll r$ the radial function behaves
\begin{equation}
  \label{Rasym}
  R_{nl}\sim \frac{e^{-\mu_{\rm eff} r}}{r} \,,
\end{equation}
where we introduced the {\it effective mass}
\begin{equation}
\label{mueff}
\mu_{\rm eff} = \sqrt{\mu^2 -\omega^2}= \sqrt{\mu^2 -q^2\Phi_H^2} \,.
\end{equation}
Therefore we assume $\mu^2\geq \omega^2= \frac{q^2Q^2}{r_H^2}$\footnote{If one allows the existence of configurations with
  $\mu^2< \omega^2$, then asymptotically $R_{nl}\sim \frac{\pm e^{i ||\mu_{\rm eff}|| r}}{r}$. Therefore, the
radial gradients and the scalar-field potential would behave asymptotically as $\sim 1/r^2$ and thus
energy-momentum tensor would behave asymptotically in this way too. As a consequence if one takes into account
the backreaction of the field into the spacetime, this kind of configurations would not lead to an asymptotically flat
spacetime, as the Komar mass would diverge asymptotically as $\sim r$.}. In Secs.~\ref{sec:extremalcase} and
\ref{sec:extremalmuq} we examine solutions within the background of an extremal RNBH
for which the strict equality $\mu^2=\omega^2= q^2$
is considered in our attempt to recover the solutions reported in~\cite{DegolladoHerdeiro2013}.

Furthermore, for the bound state solutions to be physically meaningful we impose regularity conditions
on the scalar field $\Psi(t, r, \theta, \varphi)$ at the BH horizon $r_H$. In particular,
we impose that the field and its derivatives are bounded at the horizon. More specifically,
$R_{nl}(r)$, $R^{\prime}_{nl}(r)$ and $R^{\prime\prime}_{nl}(r)$ have finite values at $r = r_H$, where
{\it primes} indicate the derivative with respect to the radial coordinate. Thus,
assuming that $R^{\prime\prime}_{nl}(r_H)$ is bounded in Eq.~(\ref{radialE}),
the regularity condition for $R'_{nl}(r_H)$ in the {\it subextremal} case ($|Q| < M$) turns out to be
\begin{equation}
\label{CNE1}
R^{\prime}_{nl}(r_H) = \left[ \frac{l(l + 1) + \mu^{2}r_H^{2}}{2\left(r_H - M\right)}\right] R_{nl}\left(r_H\right)\;.
\end{equation}
The value $R_{nl}\left(r_H\right)$ is {\it a priori} arbitrary, and we can choose for instance,
$R_{nl}\left(r_H\right) \equiv 1$.   
To find $R^{\prime\prime}_{nl}(r_H)$ we need to differentiate Eq.~(\ref{radialE}) one more time and demand that $R^{\prime\prime\prime}_{nl}(r_H)$ is 
bounded. We find
\begin{eqnarray}
\label{CNE2}
\nonumber R^{\prime\prime}_{nl}(r_H) = && \left[\frac{4\mu^{2}r_H\left(r_H - M\right) - q^{2}Q^{2}}{8\left(r_H - M\right)^{2}}\right] R_{nl}\left(r_H\right)\\
& + & \left[\frac{l(l + 1) + \mu^{2}r_H^{2} - 2}{4\left(r_H - M\right)}\right] R^{\prime}_{nl}\left(r_H\right)\;.
\end{eqnarray}
We see that the radial derivatives in Eqs.~(\ref{CNE1}) and (\ref{CNE2}) are finite on the horizon
$r_H= M + \sqrt{M^{2} - Q^{2}}$. However, we appreciate that in the extremal RNBH 
one requires a separate analysis as in this case these derivatives blow up when
$r_H^{\rm ext}=M=|Q|$ (see Sec.~\ref{sec:extremalcase}).

Similar regularity conditions are obtained when analyzing clouds in the background of a
subextremal Kerr-Newman black hole \cite{Garciaetal}, and when considering the non-rotating limit $a = 0$, we checked that they
reduce to the conditions (\ref{CNE1}) and (\ref{CNE2}).

We performed a numerical analysis to solve the radial Eq.(\ref{radialE}) under the regularity conditions
(\ref{CNE1}) and (\ref{CNE2}), and find that the only solution that vanishes asymptotically is
the trivial one $R_{nl}=0$. Given that the background spacetime is spherically symmetric one would expect cloud
solutions respecting symmetry. Nonetheless, spherically symmetric $(l=0)$ non-trivial cloud solutions were not
found either.



\section{The extremal RNBH and regularity conditions}
\label{sec:extremalcase}
Let us now focus on the extremal RNBH associated with $r_H^{\rm ext} = |Q| = M$, with metric
\begin{eqnarray}
\label{RNextremal} 
\nonumber ds^2 = &-&\frac{\left(r - M\right)^2}{r^2}dt^{2} + \frac{r^2}{\left(r - M\right)^2}dr^{2} \\
 &+& r^{2}d\theta^{2} + r^{2}\sin^{2}\theta d\varphi^{2}\,.
\end{eqnarray}
 
From Eq.~(\ref{fluxcond2}) the critical frequency for the extremal case is
\begin{equation}
\label{omegaext}
\omega_c = \frac{qQ}{M}= \frac{qQ}{|Q|}= q\,{\rm sign(Q)}  \;.
\end{equation}
For instance, $\omega_c=q$ when choosing $Q>0$, and thus $\Phi_H=1$.

The radial function $R^{\rm ext}_{nl}$ obey the equation 

\begin{equation}
\label{radialExt}
\frac{d}{dr}\left(\Delta_{\rm ext}\frac{dR^{\rm ext}_{nl}}{dr}\right) + \left[\frac{\mathcal{H}_{\rm ext}^2}{\Delta_{\rm ext}} -
  \left(K_{l}^{\rm ext} + \mu^2r^2 \right)\right]R^{\rm ext}_{nl} = 0 \;, 
\end{equation}
where
\begin{eqnarray}
 \Delta_{\rm ext} &=& \left(r - M\right)^{2} \;,\\
\label{Hext}
\mathcal{H}_{\rm ext} &\equiv& \omega r^{2} - qQr = qr\left(r - Q\right) = qr\left(r - M\right)\;.
\end{eqnarray}
Like in the subextremal scenario, $\mathcal{H}_{\rm ext}$ also vanishes at the horizon $r=r_H = M$. 

Assuming again boundedness of the field and the radial derivatives at the horizon we find the
following regularity conditions,
\begin{equation}
\label{CNExt1}
R^{\rm ext\prime}_{nl}(M) = \left[\frac{2M\left(q^{2} - \mu^{2}\right)}{2 + M^{2}\left(q^{2} - \mu^{2}\right) -
    K^{\rm ext}_{l}}\right] R^{\rm ext}_{nl}(M)\;.
\end{equation}
\begin{eqnarray}
\label{CNExt2}
\nonumber R^{\rm ext\prime\prime}_{nl}(M) &=& - \left[\frac{2\left(q^{2} - \mu^{2}\right)}{6 + M^{2}\left(q^{2} - \mu^{2}\right) -
    K^{\rm ext}_{l}}\right] R^{\rm ext}_{nl}\left(M\right)\\
&-& \left[\frac{4M\left(q^{2} - \mu^{2}\right)}{6 + M^{2}\left(q^{2} - \mu^{2}\right) - K^{\rm ext}_{l}}\right]
R^{\rm ext\prime}_{nl}\left(M\right)\;.
\end{eqnarray}
Notice that Eqs. (\ref{CNExt1}) and (\ref{CNExt2}) are finite at the horizon $r_H^{\rm ext} = M$ assuming $|R^{\rm ext}_{nl}(M)| <\infty$.

The fact that $\Delta_{\rm ext} = (r - M)^2$ and its derivative $\Delta'_{\rm ext} = 2(r - M)$ vanish at
the horizon lead to the following form for the separation constants $K_{l}^{\rm ext}$ in the extremal case
while assuming $R_{nl}^{\rm ext}(M) \neq 0$, otherwise the solution becomes the trivial one by virtue of Eqs.(\ref{CNExt1}) and (\ref{CNExt2}):
\begin{equation}
\label{constantsext}
K_{l}^{\rm ext} = \left(q^{2} - \mu^{2}\right)r_H^{2}.
\end{equation}
These separation constants are different from those given by Eq.(\ref{constants}),
which are associated with the values required by the spherical harmonics (i.e. the angular part of the field)
to be well behaved. We thus face a similar consistency problem that we found when analyzing clouds
in the extremal Kerr background \cite{Garcia2}. In particular, the separation constants given by (\ref{constantsext}) are not even
integers and are non-positive since $q^{2} \leq \mu^{2}$. Thus, both types of the separation constants
match
\begin{equation}
l(l + 1) = (q^{2} - \mu^{2})M^{2}\;,
\end{equation}
only if $\mu^2=q^2$, and therefore, only if $l=0$. The condition $\mu = |q|=|\omega|$ (extremal test field)
is precisely the one imposed by Degollado \& Herdeiro \cite{DegolladoHerdeiro2013} to report non-trivial cloud solutions.
Nevertheless, from the above considerations not only the angular dependency is absent, but 
also non-trivial spherically symmetric solutions are absent as well since
the regularity conditions (\ref{CNExt1}) and (\ref{CNExt2}) reduce to
\begin{equation}
R^{\rm ext\prime}_{nl}(M) = R^{\rm ext\prime\prime}_{nl}(M) = 0,
\end{equation}
and the only possible radial regular solution is
\begin{equation}
R^{\rm ext}_{nl}(r) \equiv {\rm const}.
\end{equation}
In particular, choosing ${\rm const} = 0$, for the solution to vanish asymptotically, we are led to the trivial solution
\begin{equation}
\Psi(t, r, \theta, \varphi) \equiv 0,
\end{equation}
which indicates that it is not possible to find non-trivial {\it scalar clouds} or bound states in the extremal RNBH under the scenario proposed
in \cite{DegolladoHerdeiro2013}. 

This conclusion has, however, some caveats. Here we assumed regularity in the radial derivatives
for the field. This is a sufficient condition leading to a well behaved scalars formed from the ``kinetic'' term
$g^{ab} (D_a \Psi)^* D_b \Psi$, but it is not necessary {\it a priori}. For instance,
given that in the subextremal scenario $R_H$ is a free parameter, one
  could choose $R_H= (r_H - M)^{\beta} B$ where $B$ is a constant, that we can take $B=1$. If $0<\beta<1$ then
  from Eq.(\ref{CNE1}) we see that in the extremal limit $R_H\rightarrow 0$ and $R'_H\rightarrow \infty$ and in this
  way the trivial solution is avoided. 
  Moreover, in such kinetic term appears $g^{rr} (\Psi^\prime)^2$, and since
  $g^{rr}= (r-M)^2/r^2$ in the extremal RNBH, in principle, one could afford a divergence at $r=M$ in the radial derivative
of the type $\Psi^\prime \sim (r-M)^{-\alpha}$
with $0<\alpha < 1$, while still allowing for the kinetic scalars to be bounded at the extremal horizon.
This happens in the extremal Kerr cloud solutions found by Hod~\cite{Hod2012}, where the radial functions vanish at
the horizon but the derivatives blow-up there. But even with this caveat in mind,
in the next section we proof a no-hair theorem that excludes this possibility as well.
In fact, we have verified that if we propose the ansatz $R^{\rm ext}_{nl}(r) = (r-M)^\alpha L(r)$ 
  for solving Eq.(\ref{radialExt}) such that $R^{\rm ext}_{nl}(r_H)=0$ and $L(r_H)\neq 0$, then
  we find an algebraic equation for $\alpha$ that depends implicitly on $M$
  and a differential equation for
  $L(r)$ together with its regularity conditions\footnote{Originally, we implemented this technique for the
    extremal Kerr BH in collaboration with P. Grandcl\'ement and E. Gourgoulhon that we plan to report in a
    forthcoming report. In that scenario, the resulting regular solutions for the equivalent of $L(r)$ have
    $0< \alpha<1$ and hence, the radial solutions vanish at the horizon, and the kinetic term turns out
    to be also bounded there, as we have discussed in the main text for the extremal RNBH,
    despite the divergent behavior of $dR^{\rm ext}_{nl}/dr$ at $r=M$.}. However, we find that the only well behaved solutions for $L(r)$ are those with $\alpha<0$,
  leading to bad-behaved solutions for $R^{\rm ext}_{nl}(r)$ at the horizon. Thus, the only possibility for a regular
  solution in the domain of outer communication of the extremal RNBH with a vanishing field asymptotically is
  the trivial solution $R^{\rm ext}_{nl}(r)\equiv 0$.  This conclusion is further supported by a no-hair theorem
  presented in the next section.



\section{No-hair theorem}
\label{sec:theorem}
As a complementary analysis we now present a more heuristic study to justify the existence (or absence) of non-trivial boson clouds in the
background of a RNBH. This analysis is based upon an integral technique developed by
Bekenstein~\cite{Bekenstein1972}, with variants provided by other authors
to prove no-hair theorems in different scalar-field theories \cite{Ayon2002,Heusler1996}, and 
which we implemented recently~\cite{Garcia,Garcia2}
to analyze the existence of non-charged clouds in the background of a Kerr BH.

Let us consider the Klein-Gordon equation for the charged boson field in the form\footnote{In \cite{Gourgoulhon} they use a Klein-Gordon equation with the equivalent form ${D}^{a}{D}_{a}\Psi = \frac{\partial \tilde{U}\left(|\Psi|^2\right)}{\partial|\Psi|^2}\Psi$.}
\begin{equation}
\label{KGgeneral}
{D}^{a}{D}_{a}\Psi = \frac{\partial \tilde{U}\left(\Psi^*\Psi\right)}{\partial \Psi^*} ,
\end{equation}
where $\tilde{U}\left(\Psi^*\Psi\right)$ \footnote{Equation (\ref{KGgeneral}) is equivalent to Eq.~(\ref{KG}) when we consider the following relation between both $U\left(\Psi^*\Psi\right) = \frac{\tilde{U}\left(\Psi^*\Psi\right)}{2}$. In \cite{Gourgoulhon} they consider a energy-momentum tensor associated with the scalar field $T_{ab} = \nabla_{(a}\Psi\nabla_{b)}\Psi^* - \frac{1}{2}g_{ab}\left[\nabla_c\Psi\nabla^c\Psi^* + \tilde{U}\left(|\Psi|^2\right)\right]$.} is the potential associated with the complex scalar field $\Psi = \phi(r, \theta)e^{-i(\omega t - m\varphi)}$.
Multiplying both sides by $\Psi^{*}$ in the last equation and integrating over a spacetime volume
$\mathcal{V}$ within the domain of outer communication of the BH we obtain
\begin{equation}
  \nonumber \int_\mathcal{V}\Psi^{*}{D}^{a}{D}_{a}\Psi
  \sqrt{-g}d^{4}x = \int_\mathcal{V}\Psi^{*}\frac{\partial \tilde{U}\left(\Psi^*\Psi\right)}{\partial \Psi^*}\sqrt{-g}d^{4}x \, .
\end{equation}
Integrating by parts the l.h.s and using the Gauss theorem, a straightforward calculation leads to
\begin{eqnarray}
&&  \nonumber \int_{\partial \mathcal{V}} \Psi^{*}s^{a}D_a\Psi dS \nonumber \\
&&  = \int_\mathcal{V}\Big[ (D^a \Psi)^* (D_a \Psi) + \Psi^{*}\frac{\partial \tilde{U}\left(\Psi^*\Psi\right)}{\partial \Psi^*}\Big]
  \sqrt{-g}d^{4}x \,.
\end{eqnarray}
The surface integral associated with the boundary $\partial \mathcal{V}$
has four contributions: one at a portion of the BH horizon, one at spatial infinity and two contributions
corresponding to integrals over two spatial hypersurfaces $\Sigma_{t_1}$ and $\Sigma_{t_2}$. The latter two cancel each other
because the spacetime is static, and the scalar-field contributions stationary, and thus, these two integrals
differ only by the normals to both hypersurfaces, which are opposite. The surface integral associated with the asymptotic
region at spatial infinity vanishes when demanding that the field $\Psi$ falls off sufficiently rapid, namely, exponentially due to the presence of a mass term, which would produce an asymptotically flat spacetime if the backreaction of the field
were taken into account. Finally, it remains the surface integral at the horizon, which is a null hypersurface, with
normal $s^a$ given by the timelike Killing field
$\xi^a= \left(\frac{\partial}{\partial t}\right)^a$ at the horizon. Therefore, $g_{ab}\xi^a \xi^b$
  vanishes at the horizon: $g_{ab}\xi^a \xi^b|_{r_H}=g_{tt}|_{r_H}= \left(1 - \frac{2M}{r} + \frac{Q^2}{r^2}\right)|_{r_H}= 0$.
  Thus, $\Psi^{*} s^a D_a\Psi|_{r_H}= -i \Psi^{*} \Psi \left(\omega + qA_t\right)|_{r_H}$.
  Assuming that $\Psi^{*} \Psi$ is bounded but finite at the
  horizon, the surface integral at $r_H$ vanishes due to the condition (\ref{fluxcond2}), $\omega= -qA_t|_{r_H}= q \Phi_H$.
We conclude
\begin{equation}
  \label{Integral}
\int_\mathcal{V}\Big[ (D^a \Psi)^* (D_a \Psi) + \Psi^{*}\frac{\partial \tilde{U}\left(\Psi^*\Psi\right)}{\partial \Psi^*}\Big]
  \sqrt{-g}d^{4}x =0 \,.
\end{equation}
The first term in the integrand corresponds to the {\it kinetic} contribution:
\begin{equation}
K= (D^a \Psi)^* (D_a \Psi)= g^{tt} (D_t \Psi)^* (D_t \Psi) + g^{ij} (D_i \Psi)^* (D_j \Psi) \,,
\end{equation}
 where
\begin{eqnarray}
&&  g^{ij} (D_i \Psi)^* (D_j \Psi)= g^{ij} (\nabla_i \Psi)^* (\nabla_j \Psi)=
  g^{rr} (\nabla_r \phi)(\nabla_r \phi) \nonumber \\
  &+& g^{\theta\theta} (\nabla_\theta \phi)(\nabla_\theta \phi) +
  g^{\varphi\varphi} (\nabla_\varphi \Psi^*)(\nabla_\varphi \Psi) \nonumber \\
  &=& g^{IJ} (\nabla_I \phi) (\nabla_J \phi) + g^{\varphi\varphi} m^2 \phi^2\,,
\end{eqnarray}
which is non-negative in the domain of outer communication. Here lower-case latin indices $i,j$ run $r,\theta,\phi$, 
and we used $D_i= \nabla_j$, because $A_a$ has a component
only in the time direction, and also used the harmonic dependency of the field with respect to the angle $\varphi$
following (\ref{Psians}). Moreover, the indices $I,J$ run $r,\theta$.
The term with time derivatives in the kinetic term $K$ reads explicitly as follows,
\begin{equation}
  \label{DtDt}
g^{tt} (D_t \Psi)^* (D_t \Psi) = g^{tt} \left(\omega + qA_t\right)^2 \phi^2 \,,
\end{equation}
where we used the harmonic time dependency of the field following (\ref{Psians}).

Collecting these results, the integrand in (\ref{Integral}), reads
\begin{eqnarray}
  \label{integrand}
  && I= g^{tt} \left(\omega + qA_t\right)^2 \phi^2 +
  \Psi^{*}\frac{\partial \tilde{U}\left(\Psi^*\Psi\right)}{\partial \Psi^*}  \nonumber \\
 &+&  g^{IJ} (\nabla_I \phi) (\nabla_J \phi) + m^2 g^{\varphi\varphi}\phi^2  \,.
\end{eqnarray}

Below we present two scenarios: one analyzed by Degollado \& Herdeiro \cite{DegolladoHerdeiro2013} like in
Sec.~\ref{sec:extremalcase}, where we show that the integrand (\ref{integrand}) is not negative, and thus,
a no-hair theorem can be established, and another one presented more recently by several authors
\cite{Hong2020,Herdeiro2020,Brihaye}
where the integrand has no definite sign and thus, it is not possible to establish a no-hair theorem.

The idea is to justify in an heuristic way the existence or absence of scalar clouds around a charged,
static and spherically symmetric black hole in these two scenarios.

\subsection*{Absence of charged clouds within a RNBH}
\label{sec:extremalmuq}

We assume a RNBH where
\begin{equation}
g^{tt} = -\frac{r^{2}}{\left(r - r_H\right)\left(r - r_{-}\right) }\,.
\end{equation}

In this case, the term (\ref{DtDt}) reads
\begin{eqnarray}
  && g^{tt} \left(\omega + qA_t\right)^2 \phi^2
  = -\frac{r^2\left(\omega + qA_t\right)^2 \phi^2}{(r-r_H)(r-r_{-})} \nonumber \\
 &=& -\frac{q^2Q^2\left(r - r_H \right)\phi^2}{r_H^2(r-r_{-})}
    \end{eqnarray}
where we used the condition (\ref{fluxcond2}) for $\omega$, and the electric potential (\ref{elecpot})
like in \cite{DegolladoHerdeiro2013}.

Furthermore, we take the potential $\tilde{U}$ for a massive but free field as follows,
\begin{equation}
\tilde{U}(\Psi^*\Psi) = \mu^{2}|\Psi|^{2}\,.
\end{equation}

In this way, the integrand in (\ref{integrand}), reads
\begin{eqnarray}
  \label{integrand2}
&& I=  \left[\mu^2 - \frac{q^2Q^2\left(r - r_H \right)}{r_H^2(r-r_{-})}\right]\phi^2 \nonumber \\
  &+&  g^{IJ} (\nabla_I \phi) (\nabla_J \phi) + m^2 g^{\varphi\varphi} \phi^2 \nonumber \\
  &=& \left[\mu^2 - \frac{q^2Q^2}{r_H^2} + \frac{q^2Q^2\left(r_H - r_{-} \right)}{r_H^2(r-r_{-})}\right]\phi^2 \nonumber \\
  &+&  g^{IJ} (\nabla_I \phi) (\nabla_J \phi) + m^2 g^{\varphi\varphi} \phi^2 \nonumber \\
\end{eqnarray}
The term within the brackets is positive semi-definite (i.e. a non-negative quantity) because $\mu^2\geq \omega^2= \frac{q^2Q^2}{r_H^2}$ in order for the
scalar-field to falls of asymptotically as in (\ref{Rasym}), and also because
the third term in the brackets is not negative since $r>r_{-}$ and $r_H\geq r_{-}$ in the domain of outer
communication of the RNBH. The equalities $\omega^2=q^2$ and $r_H= r_{-}=|Q|=M$ occurs
in the extremal RNBH. The remaining terms of the integrand
$I$ in (\ref{integrand2}) are non-negative in the domain of outer communication.
Thus, for an extremal RNBH the integrand reduces to
\begin{eqnarray}
  I &=&   \left(\mu^2 - q^2\right)\phi^2 + g^{ij} (\nabla_i \Psi)^* (\nabla_j \Psi)\nonumber \\
  &=& \left(\mu^2 - q^2+ m^2 g^{\varphi\varphi}\right)\phi^2 + g^{IJ} (\nabla_I \phi) (\nabla_J \phi)\,.
\end{eqnarray}
Therefore, in general the integral (\ref{Integral}) becomes
\begin{eqnarray}
  \label{Integral2}
  &&  \int_\mathcal{V}\Big[ \left(\mu^2 -  \frac{q^2Q^2}{r_H^2}
    + m^2g^{\varphi\varphi} + \frac{q^2Q^2\left(r_H - r_{-} \right)}{r_H^2(r-r_{-})} \right)\phi^2
      \nonumber \\
&&  +  g^{IJ} (\nabla_I \phi)(\nabla_J \phi) \Big]
  \sqrt{-g}d^{4}x =0 \,.
\end{eqnarray}
So, in either scenario, the subextremal and extremal ones, the integrand in the above integral is not negative 
and thus, in general the equality in (\ref{Integral2}) holds only if the scalar-field vanishes identically,
i.e., $\phi(r,\theta)\equiv 0$ and therefore
$\Psi(t,r,\theta,\varphi)\equiv 0$. We have thus proved that non-trivial {\it regular}
charged clouds in the background of a RNBH
with a non self-interacting potential are not possible. In particular, this conclusion holds also
for the extremal scenario considered by Degollado \& Herdeiro \cite{DegolladoHerdeiro2013} where $\mu= |q|=|\omega|$,
dubbed {\it double extremal limit.} For that case the integral (\ref{Integral2}) reduces to
\begin{equation}
  \label{Integral3}
  \int_\mathcal{V}\Big[ m^2 g^{\varphi\varphi}\phi^2 +  g^{IJ} (\nabla_I \phi)(\nabla_J \phi)\Big]
  \sqrt{-g}d^{4}x =0 \,,
\end{equation}
leading to $\phi(r)\equiv 0$ if $m\neq 0$ and $\phi(r)\equiv {\rm const}$ if $m=0$.
  Nevertheless, since we demand that $\phi(r\rightarrow \infty)\rightarrow 0$ for the integral
  surface at spatial infinity to vanish, then $\phi(r)\equiv 0$ also for $m=0$.
Thus, contrary to what it is claimed in~\cite{DegolladoHerdeiro2013}, non-trivial and regular
charged clouds with $\mu= |q|=|\omega|$ cannot exist in the background of an extremal RNBH, even if $m=0$.

This conclusion is consistent with the one presented in Sec.~\ref{sec:extremalcase}, albeit more general.
For instance, in this analysis it was not necessary to impose the boundedness of the radial derivative, $R'_{nl}$,
at the horizon. What matters in this analysis is that each term in (\ref{Integral2}) is bounded
in the domain of outer communication, notably, at the horizon, in
particular $g^{IJ} (\nabla_I \phi)(\nabla_J \phi)$, namely, $g^{rr} (\partial_r\phi)^2$. Thus, $\partial_r\phi$,
or equivalently $R'_{nl}$ might diverge near the extremal horizon as $R'_{nl}\sim (r-M)^{-\alpha}$, with
$0<\alpha<1$, so that $g^{rr} (\partial_r\phi)^2= (r-M)^2 (\partial_r\phi)^2/r^2 \sim (r-M)^{2-2\alpha}/r^2
$ is bounded at $r=M$.

Finally, this analysis  also shows that the exact solution for charged clouds obtained by Hod~\cite{Hod2015}
in the extremal Kerr-Newman do not admit the static limit $a=0$, where $a$ is the Kerr parameter associated with the spin
of the black hole.


\section{Charged Q-clouds}
\label{sec:Qclouds}
\begin{figure}[h]
\begin{center}
\includegraphics[width=0.5\textwidth]{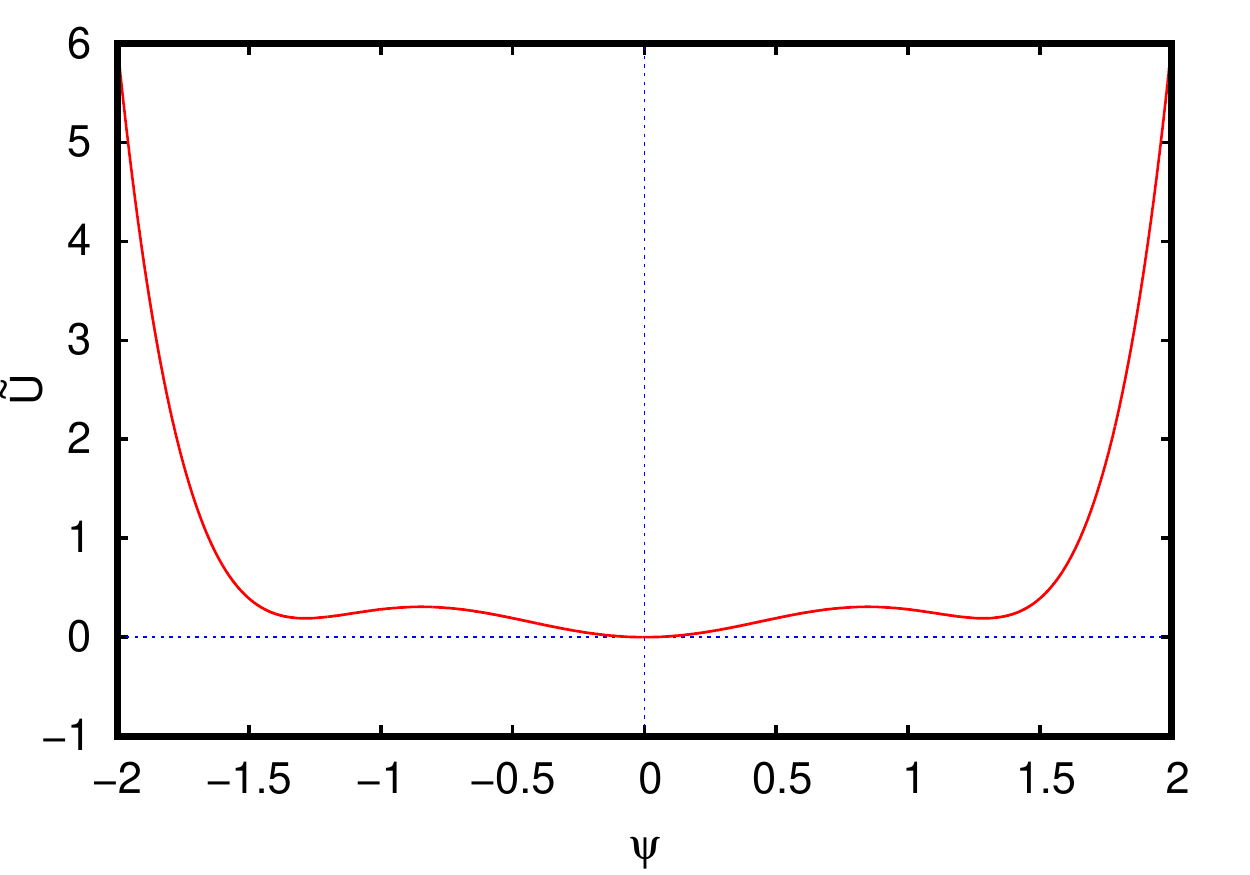}
\caption{The scalar-field potential $\tilde{U}$ (\ref{PotentialBH}) as a function of $\psi$, assuming the values
  $\mu = 1=\lambda$ and $\nu = 9/32$ as in Ref.\cite{Brihaye}.}
\label{fig:Potential}
\end{center}
\end{figure}

Several authors \cite{Hong2020,Herdeiro2020,Brihaye} analyzed the existence of
spherically symmetric scalar clouds with a self-interacting potential,
within the background of a RNBH, and also by taking into account the backreaction of the boson field into the spacetime.
For the latter case, a charged, static and spherically symmetric black hole is assumed with a
spacetime metric in the following form
\begin{eqnarray}
  \label{MetricRN}
\nonumber ds^{2} = &-& \sigma^2(r)N(r)dt^{2} + \frac{1}{N(r)}dr^2 \\
&+& r^2\left(d\theta^2 + \sin^2\theta d\varphi^2\right) \,.
\end{eqnarray}
For this kind of clouds a complex-valued and charged scalar field with no angular dependency was considered,
\begin{equation}
\Psi(t,r) = \psi(r)e^{-i\tilde{\omega} t},
\end{equation}
submitted to a potential $\tilde{U}$
\begin{eqnarray}
  \label{PotentialBH}
  \tilde{U}(\Psi^*\Psi) &=& \mu^{2}\Psi^*\Psi - \lambda\left(\Psi^*\Psi\right)^{2} + \nu\left(\Psi^*\Psi\right)^{3} \nonumber \\
  &=& \mu^{2}\psi^2 - \lambda \psi^4 + \nu \psi^6 \,,
\end{eqnarray}
where $\lambda$ and $\nu$ are positive real numbers, with
$\nu> \lambda^2/4\mu^2$ for $\tilde{U}$ to be a true vacuum at $\psi=0$~\cite{Hong2020}.
Figure~\ref{fig:Potential} depicts this potential. 
For the RNBH, and the test-field approximation, which is the only problem that we analyze here, $\sigma(r)\equiv 1$, and 
\begin{equation}
\label{FuctN}
N(r) = -g_{tt} = 1 - \frac{2M}{r} + \frac{Q^2}{r^2} \,,
\end{equation}
given by (\ref{RN}).

At this point it is important to remark that the theory considered so far is invariant with respect
to a {\it local} phase transformation in the field $\Psi$. This local transformation is compensated by the gauge transformation
in the electromagnetic potential. In particular, the theory is invariant with respect to a transformation
\begin{eqnarray}
\omega &=& {\tilde \omega} + \zeta \,,\\
 A_t &=& {\tilde A}_t - \zeta/q \,, 
\end{eqnarray}
where $\zeta$ is a constant. This can be appreciated by a direct substitution in the full-fledged set of equations
\cite{Herdeiro2020}. Nonetheless, this invariance is apparent from the field equations of
the full theory since $A_t$ and $\omega$ appears always in a combination $\omega + q A_t$, which remains invariant
under the above transformation, and also because the radial derivatives
  for $A_t$ are unaffected by this shift. As a consequence, one can use a gauge different from the one of previous sections
  where
\begin{eqnarray}
\label{omegaB}\tilde{\omega} &=& 0 \,,\\ 
\label{PotentialE}{\tilde A}_t &=& V(r) = \frac{Q}{r_H} - \frac{Q}{r}\,.
\end{eqnarray}
This gauge was employed in \cite{Brihaye}, and previously in~\cite{Herdeiro2020}. Notice that
under this gauge, the electric potential $V(r)$ vanishes at the horizon, but asymptotically it takes a non-zero value.
In our case, whether one uses this or the original gauge where $\omega\neq 0$, it is irrelevant since our treatment is
gauge invariant. Therefore the integrand (\ref{integrand}) remains the same. In particular the integrand
(\ref{integrand2}) takes the same form, except that we have to replace the mass term for the
corresponding term obtained from the potential (\ref{PotentialBH}), and also taking
$m\equiv 0$, and $\nabla_\theta \phi=0$, since in this scenario we are assuming only a time and radial dependency in the field.
Thus, the integral (\ref{Integral2}) becomes

\begin{eqnarray}
  \label{Integral3}
  &&  \int_\mathcal{V}\Big[
    \left(\mu^2 - 2\lambda\psi^2 + 3\nu\psi^4 -  \frac{q^2Q^2}{r_H^2}
    + \frac{q^2Q^2\left(r_H - r_{-} \right)}{r_H^2(r-r_{-})} \right)\psi^2
      \nonumber \\
&&  +  g^{rr} (\psi')^2 \Big]
  \sqrt{-g}d^{4}x =0 \,,
\end{eqnarray}
where we used
\begin{eqnarray}
\nonumber \Psi^*\frac{\partial \tilde{U}\left(\Psi^*\Psi\right)}{\partial \Psi^*} &=& \Psi^*\left[\mu^2\Psi - 2\lambda\left(\Psi^*\Psi\right)\Psi + 3\nu\left(\Psi^*\Psi\right)^2\Psi\right] \\
& = & \mu^2 \psi^2 - 2 \lambda \psi^4 + 3\nu \psi^6\,.
\end{eqnarray}

Unlike the scenario with no self-interaction, the integrand in (\ref{Integral3}) has not a definite sign due to the
presence of the self-interaction terms, notably,
\begin{equation}
  \label{Lambda}
\Lambda (\psi) \equiv \psi^2\left(\mu^2_{\rm eff,\infty} - 2\lambda\psi^2 + 3\nu\psi^4\right) \,,
\end{equation}
where
 $\mu^2_{\rm eff,\infty}= \mu^2 - \frac{q^2Q^2}{r_H^2}$
corresponds to the mass introduced in (\ref{mueff}) for the non-self-interacting model, and like in that model,
$\mu^2_{\rm eff,\infty}\geq 0$ so that the boson field also behaves asymptotically like in (\ref{Rasym}).

In this way the integral (\ref{Integral3}) reads 
\begin{equation}
  \label{Integral4}
  \int_\mathcal{V}\Big[\Lambda(\psi) + \frac{q^2Q^2\left(r_H - r_- \right)}{r_H^2(r-r_{-})}\psi^2
    +  g^{rr} (\psi')^2 \Big]
  \sqrt{-g}d^{4}x =0 \,,
\end{equation}
or even
\begin{equation}
  \label{Integral5}
 \int_\mathcal{V}\Big[
 \mu^2_{\rm eff}(r) \psi^2 + \Sigma(\psi) \psi^2  +  g^{rr} (\psi')^2 \Big]
  \sqrt{-g}d^{4}x =0 \,,
\end{equation}
where
\begin{equation}
  \label{Sigma}
\Sigma(\psi) \equiv \psi^2\left(- 2\lambda + 3\nu\psi^2\right) \,,
\end{equation}
  \begin{equation}
    \label{mueffr}
  \mu^2_{\rm eff}(r) \equiv \mu^2_{\rm eff,\infty} + \frac{q^2Q^2\left(r_H - r_{-} \right)}{r_H^2(r-r_{-})}\,,
\end{equation}
is a position-dependent effective squared mass.

\begin{figure}[h]
\begin{center}
\includegraphics[width=0.5\textwidth]{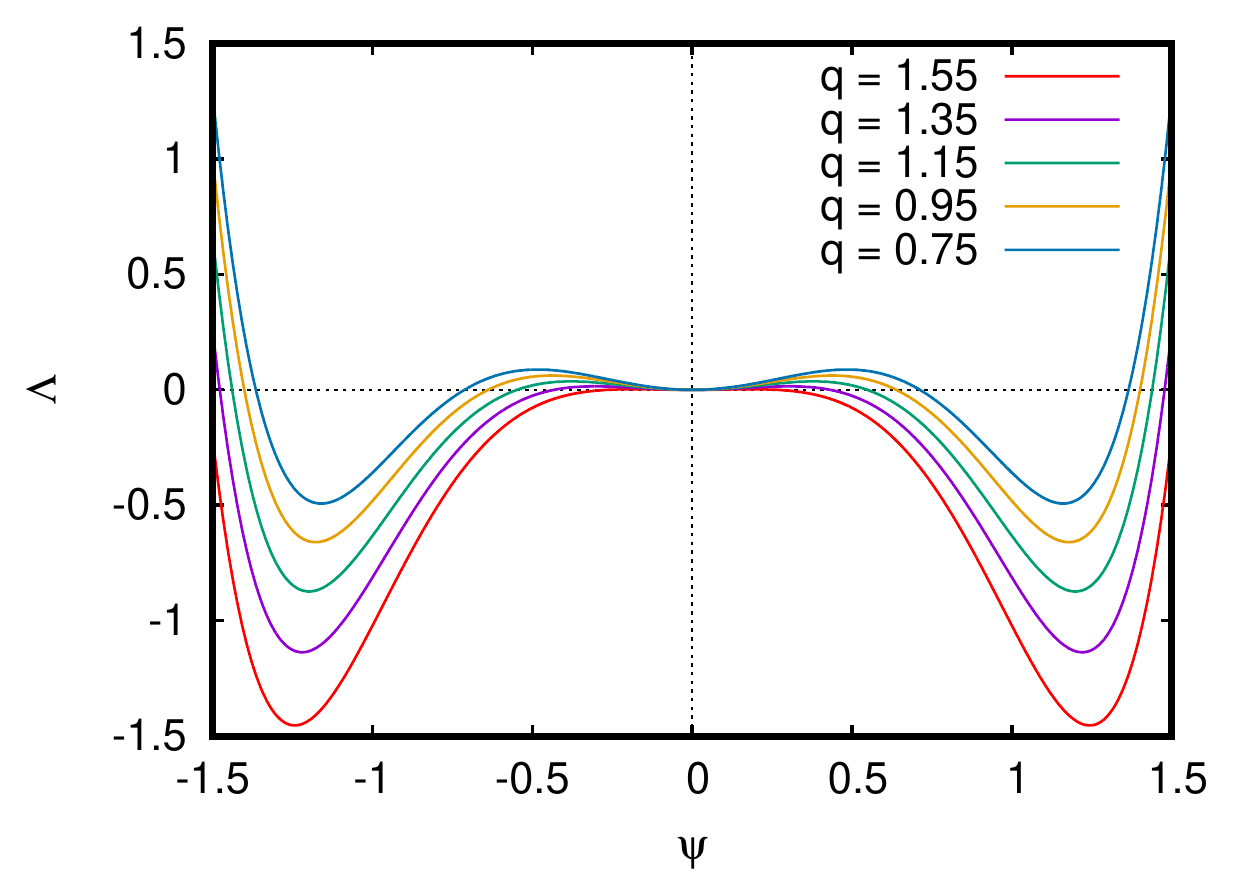}
\caption{The function $\Lambda(\psi)$ (\ref{Lambda}) for different values of $q$, taking $\mu = 1 =\lambda$ and $\nu = 9/32$ as in Fig.\ref{fig:Potential}. Here we assume $Q = 0.09$ and $r_H = 0.15$. The minima are located at $\psi_{\rm \pm}^{\rm \Lambda}$ where $\Lambda(\psi)$ is negative.}
\label{fig:KineticTRN}
\end{center}
\end{figure}

\begin{figure}[h]
\begin{center}
\includegraphics[width=0.5\textwidth]{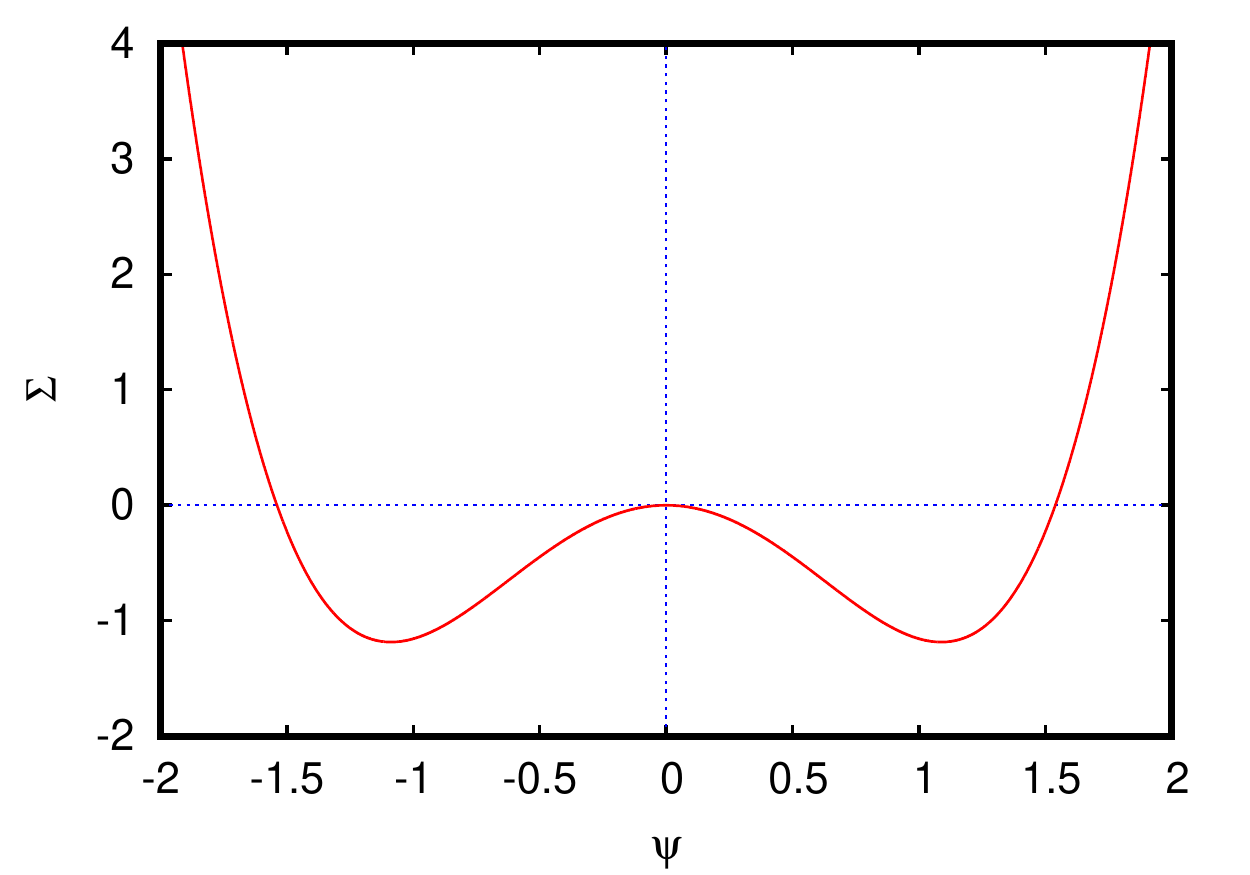}
  \caption{The function $\Sigma(\psi)$ (\ref{Sigma}) taking $\mu = 1=\lambda$ and $\nu = 9/32$
    as in Fig.\ref{fig:Potential}. The minima are located
at $\psi^\Sigma_{\pm} \approx \pm 1.0887$ where $\Sigma(\psi)$ is negative.}
\label{fig:Sigma}
\end{center}
\end{figure}

The integrands in (\ref{Integral4}) or (\ref{Integral5}) are not positive semidefinite (i.e. they can be negative)
due to the presence of $\Lambda (\psi)$ and $\Sigma (\psi)$, respectively. The terms containing the charges and the
kinetic term (the one with the radial derivative) are positive semidefinite for $r\geq r_H$. In particular,
$\Lambda$ is negative at the two minima corresponding
to $\psi_{\rm \pm}^{\rm \Lambda}=\pm \sqrt{\frac{2\lambda}{9\nu}\left[1+\sqrt{1-\frac{9\nu \mu^2_{\rm eff,\infty}}{4\lambda^2}}\right]}$
and $\Sigma$ is also negative at the two minima $\psi_{\pm}^\Sigma= \pm \sqrt{\frac{\lambda}{3\nu}}$. Thus, the integrand in
(\ref{Integral3}) has no definite sign. Figures~\ref{fig:KineticTRN} and \ref{fig:Sigma}
depict $\Lambda$, and $\Sigma$, respectively, for values of the parameters used in \cite{Brihaye},
showing that both quantities can be negative, namely, at the two minima\footnote{The specific values of
  the parameters provided in this section amount to $M,r$ and $Q$ given in units of $1/\mu$, $q$ in units
  of $\mu$, and $\lambda$ and $\nu$ in units of $\mu^{1/2}$, while $\psi$ is dimensionless (cf. Ref.~\cite{Hong2020}).}.
As a consequence, a no-hair theorem cannot be established in this case.
Not only that theorem cannot be established for this kind of self-interacting scalar field potential,
but, as we stressed before, several authors have showed that the presence of this kind of potential allows for the existence of non-trivial
charged clouds, termed {\it Q-clouds}~\cite{Hong2020,Herdeiro2020,Brihaye}
\footnote{At first sight, it is puzzling that in Ref.~\cite{Mayo1996} a no-hair theorem for a theory similar to the
  one presented in this section was established. That is, a static, spherically
  symmetric, asymptotically flat and charged subextremal black hole within Einstein's general relativity cannot support
  a non-trivial, regular and charged complex-valued scalar field endowed with a positive semidefinite scalar-field
  potential. However, in the proof of this theorem, oddly enough, the mass term associated with the scalar field
  is not taken into account, and as remarked in \cite{Hong2020b,Herdeiro2020} it is precisely the mass term that allows one
  to avoid such theorem.}. As argued in \cite{Herdeiro2020,Brihaye}
these clouds can exist even in a Schwarzschild background, but in the presence of a test electric field.

\begin{figure}[h]
\begin{center}
\includegraphics[width=0.5\textwidth]{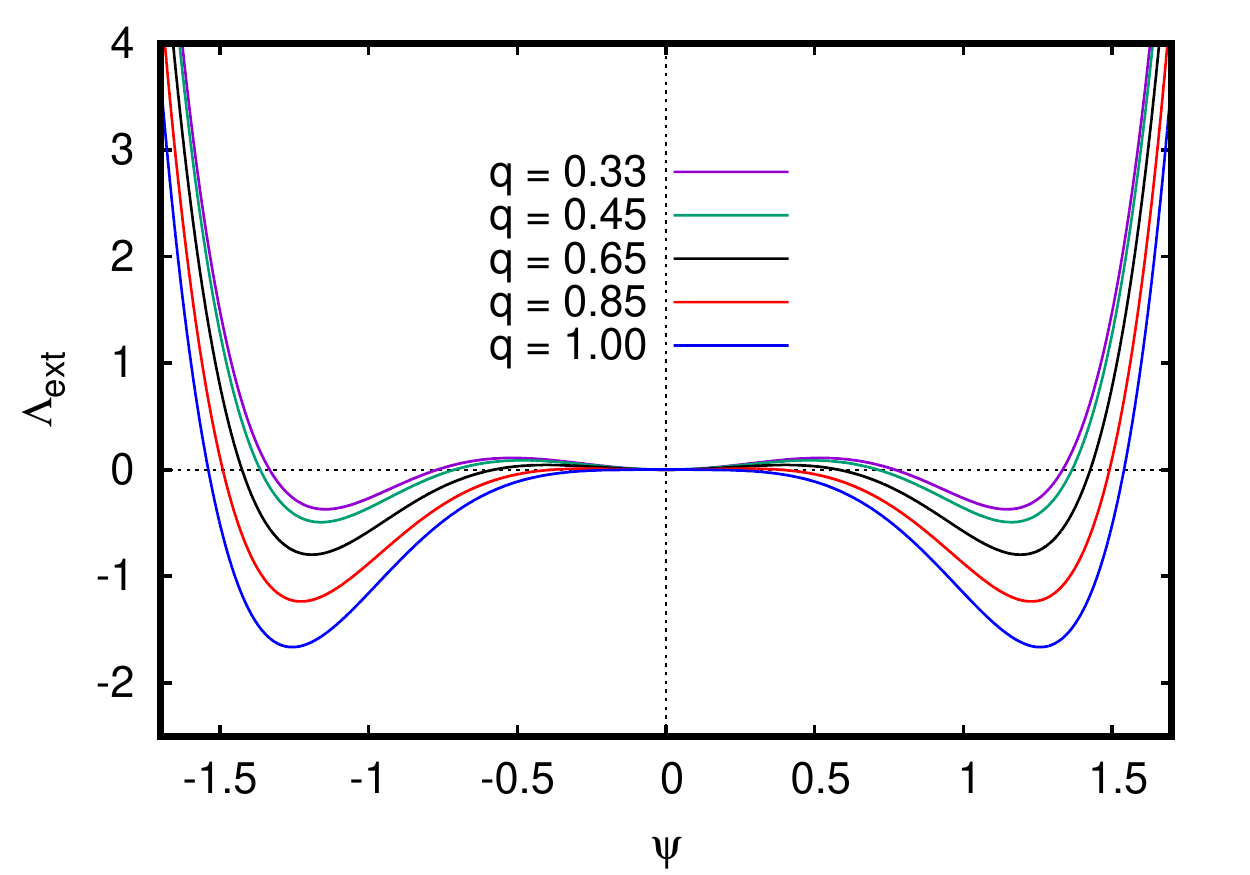}
\caption{The function $\Lambda_{\rm ext}(\psi)$, taking $\mu=1=\lambda$ and $\nu = 9/32$ for different values of $q$
  in the extremal Reissner-Nordstrom scenario.}\label{fig:Extremal}
\end{center}
\end{figure}

For the extremal RNBH $(r_H^{\rm ext} = M=|Q|)$ the integral (\ref{Integral5}) keeps the same form,
except that $\mu^2_{\rm eff}(r)=\mu^2_{\rm eff,\infty, ext}= \mu^2 -q^2$ is a non-negative constant, and
\begin{equation}
N(r) = \frac{(r - M)^2}{r^2}= g^{rr}, \qquad \sigma^2(r) = 1.
\end{equation}
More specifically the integrand in (\ref{Integral5}) reduces to
\begin{equation}\label{IntExt}
I = \left(\mu^2 - q^2 - 2\lambda\psi^2 + 3\nu\psi^4\right)\psi^2 + \frac{\left(r - M\right)^2}{r^2}\left(\partial_r\psi\right)^2\;,
\end{equation}
which does not have a definite sign. Therefore, one cannot establish a no-hair theorem either in the extremal scenario.
Figure~\ref{fig:Extremal} depicts the term
\begin{equation}
  \label{Gamma}
  \Lambda_{\rm ext}(\psi)\equiv \left(\mu^2 - q^2 - 2\lambda\psi^2 + 3\nu\psi^4\right)\psi^2\,,
\end{equation}
that appears in the integrand (\ref{IntExt}). We can rewrite this  function  as
  \begin{equation}
    \label{lambdaext}
\Lambda_{\rm ext}(\psi) = \left[3\nu\left(\psi^2 - \frac{\lambda}{3\nu}\right)^2 + \left(\mu^2_{\rm eff,\infty,\rm ext} - \frac{\lambda^2}{3\nu}\right)\right]\psi^2.
\end{equation}
which indicates that when $\mu^2_{\rm eff,\infty,\rm ext} \geqslant \lambda^2/3\nu$ the integrand of the integral (\ref{Integral5})
is positive semidefinite, in which case, the only possible Q-cloud solution are the trivial ones
$\psi(r)\equiv 0$ and $\psi(r)=\pm\sqrt{\frac{\lambda}{3\nu}}$ when $\mu^2_{\rm eff,\infty,\rm ext} = \lambda^2/3\nu$
or $\psi(r)\equiv 0$ when $\mu^2_{\rm eff,\infty,\rm ext} > \lambda^2/3\nu$. These are trivial solutions
of Eq.(\ref{RadialESI}) (see below) when $M=Q$ which correspond to the three minima
($\psi=\pm\sqrt{\frac{\lambda}{3\nu}}$, $\psi=0$) of the potential
which is introduced below in Eq.(\ref{Ueff}), and when assuming the
extremal case (denoted $U_{\rm eff}^{\rm ext}$ in the main text). This minima correspond also to the zeros of
$\Lambda_{\rm ext}(\psi)$. Notice that the two minima $\psi_{\pm}^\Lambda$ associated with $\Lambda(\psi)$
{\it degenerate} in the extremal case when $\mu^2_{\rm eff,\infty,\rm ext} = \lambda^2/3\nu$ and become
two zeros of $\Lambda_{\rm ext}(\psi)$ (cf. the discussion at the end of Sec.~\ref{sec:subQclouds}). Notwithstanding, as we show below, there exist non-trivial solutions in the near extremal scenario when $\mu^2_{\rm eff,\infty,\rm ext} < \lambda^2/3\nu$.

\subsection{Subextremal Q-clouds solutions}
\label{sec:subQclouds}
In order to find Q-cloud solutions we solve numerically
the radial equation associated with the scalar field $\psi(r)$:
  \begin{eqnarray}
    \label{RadialESI}
    \nonumber && N^2(r) \psi'' + \left[\frac{2}{r}N(r) + N'(r)\right]N(r)
    \psi' + \left(\tilde{\omega} + q\tilde{A}_t\right)^2\psi \\
    && = \frac{1}{2}\frac{\partial \tilde{U}}{\partial \psi} N(r)
=    \left(\mu^2\psi - 2\lambda \psi^3 + 3\nu\psi^5\right)N(r)   \,,
\end{eqnarray}
in the background of a subextremal RNBH, where $N(r)$ is given by (\ref{FuctN}), $\tilde{\omega}$ and $\tilde{A}_t$ by (\ref{omegaB}) and (\ref{PotentialE}), respectively.
Equation (\ref{RadialESI}) is solved by implementing the following regularity conditions for first and second derivatives at
the horizon $r_H$: 
\begin{eqnarray}
  \label{RegCond1}
  \psi'_H &=& \frac{1}{2N'(r)} \frac{\partial {\tilde U}}{\partial \psi} \Big{|}_{r=r_H} \nonumber \\
&=&\frac{r_H^3 \left(\mu^2\psi_H - 2\lambda \psi^3_H + 3\nu\psi^5_H\right)}{\left(r_H^2 - Q^2\right)} \,,
\end{eqnarray}
\begin{eqnarray}
\label{RegCond2}
\psi''_H &=& \frac{\psi'_H}{4N'(r_H)}\left[\frac{\partial^2\tilde{U}}{\partial\psi^2}\Big{|}_{r=r_H} - \frac{4}{r_H}N'(r_H) - 3N''(r_H)\right] \nonumber \\
  &-& \frac{q^2Q^2\psi_H}{2\left[r^2_HN'(r_H)\right]^2} + \frac{N''(r_H)}{2\left[2N'(r_H)\right]^2}\frac{\partial{\tilde U}}{\partial \psi} \Big{|}_{r=r_H}\nonumber \\
  &=& \frac{\psi_H'}{4\left[r_H^2 - Q^2\right]}\Big[2\left(\mu^2 - 6\lambda\psi^2_H + 15\nu\psi^4_H\right)r_H^3 \nonumber \\
  &+& 2r_H - \frac{8Q^2}{r_H}\Big] - \frac{q^2Q^2r_H^2\psi_H}{2\left[r_H^2 - Q^2\right]^2}\nonumber \\
  &-& \frac{\left(\mu^2\psi_H - 2\lambda \psi^3_H + 3\nu\psi^5_H\right)\left(r_H^2 - 2Q^2\right)r_H^2}{2\left[r_H^2 - Q^2\right]^2}\;,
\end{eqnarray}
where $\psi_H \equiv \psi(r_H)$, $\psi'_H \equiv \psi'(r_H)$ and $\psi''_H \equiv \psi''(r_H)$.

Equation (\ref{RadialESI}) is solved numerically given the parameters $\lambda$, $\nu$, $\mu$ and $q$, and fixing the value
of the horizon $r_H$ and the charge $Q$ of the RNBH. The values for $\psi'_H$ and $\psi''_H$ are determined once the specific
value for $\psi_H$ is provided. This value is found by a \textit{shooting method} such that the field $\psi(r)$ vanishes
asymptotically. At this point, it is important to stress that the field $\psi(r)$ is indeed submitted to an effective
potential that in the asymptotic region takes the form [cf. Eq.(\ref{RadialESI})]
\begin{equation}
\label{Ueff}
U_{\rm eff} = \mu_{\rm eff,\infty}^2\psi^2 - \lambda\psi^4 + \nu\psi^6,
\end{equation}
where $\mu_{\rm eff,\infty}^2 = \mu^2 - q^2Q^2/r_H^2$.

\begin{figure}[h]
\begin{center}
\includegraphics[width=0.5\textwidth]{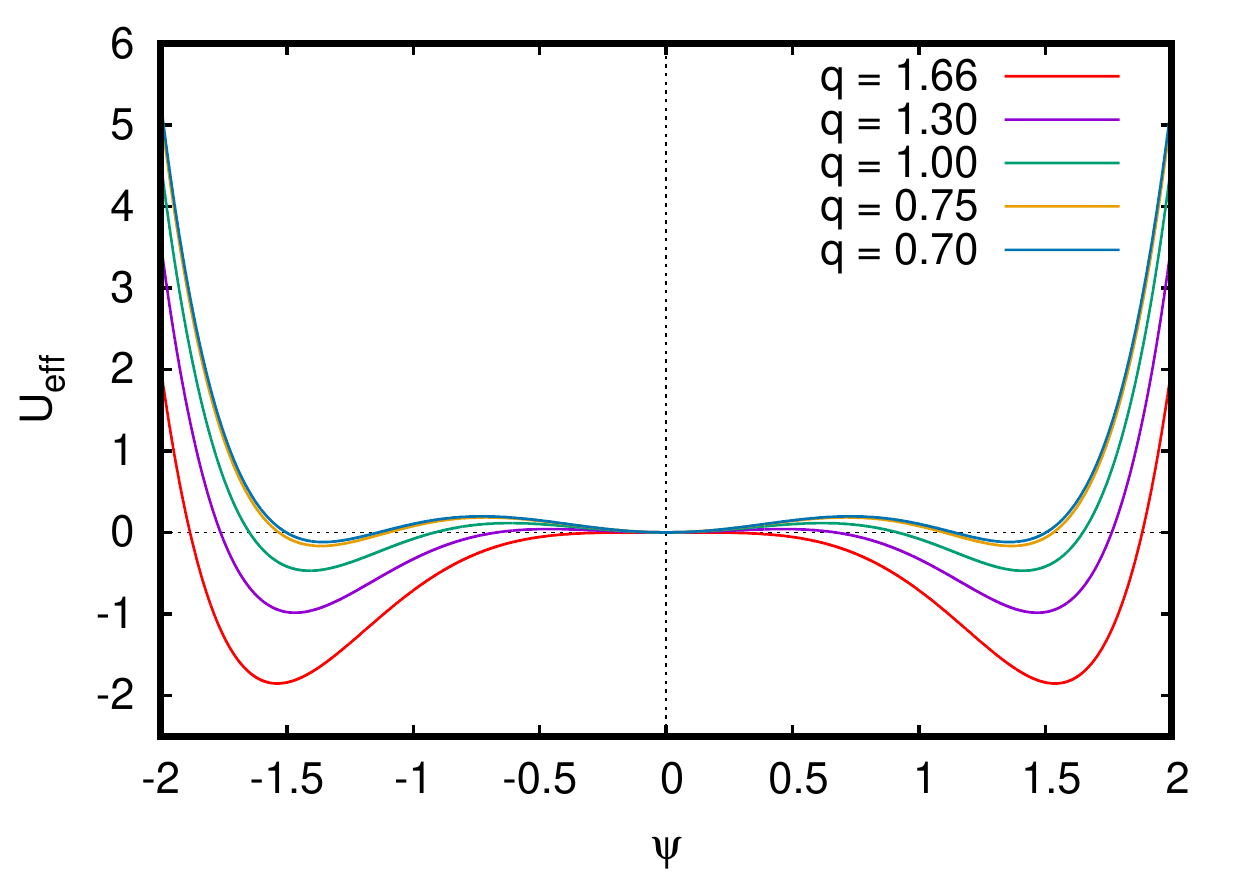}
\caption{The effective potential $U_{\rm eff}$ (\ref{Ueff}) taking $\mu = 1 = \lambda$, $\nu = 9/32$,
  $Q = 0.09$, and  $r_H = 0.15$, for different values of $q$.}\label{fig:Potentialeff}
\end{center}
\end{figure}

\begin{figure}[h]
\begin{center}
\includegraphics[width=0.5\textwidth]{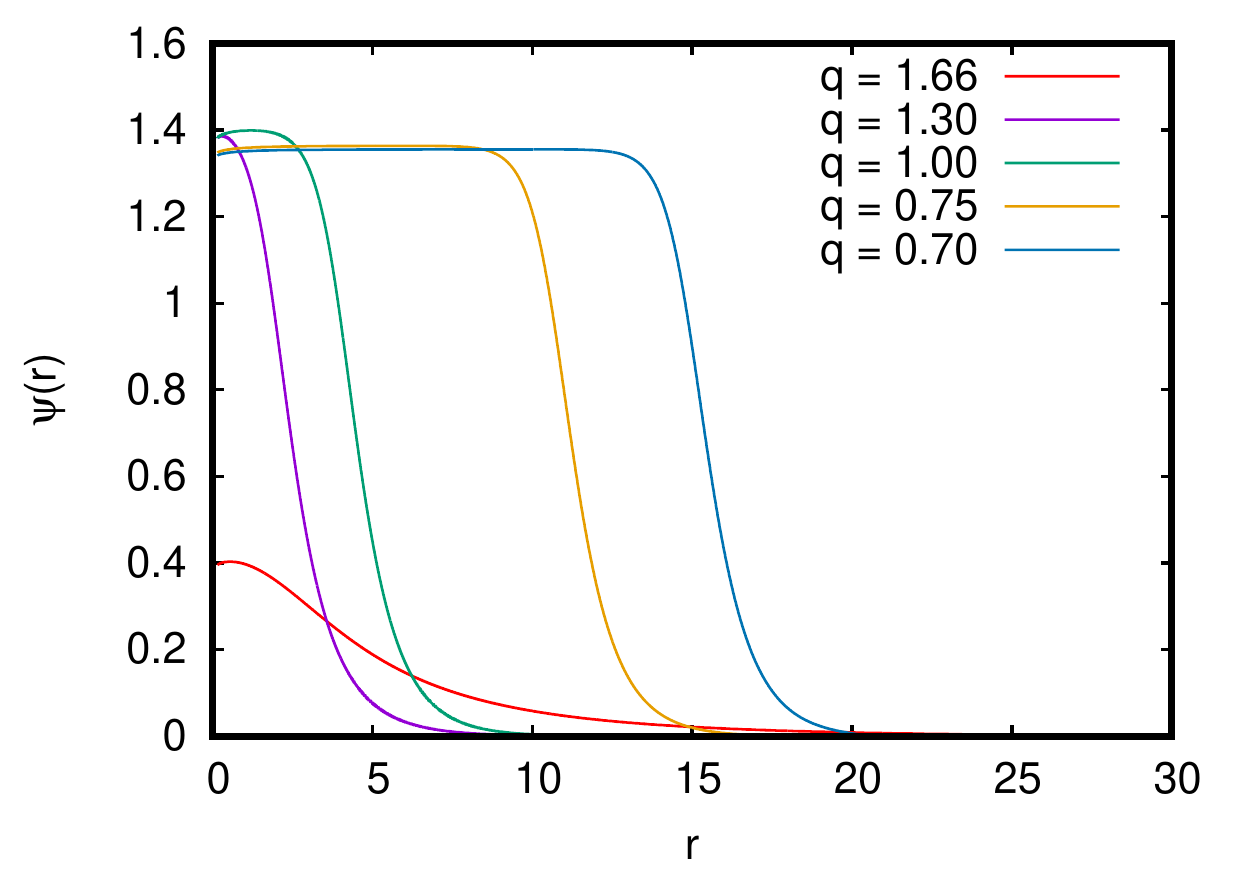}
\caption{Q-cloud solutions $\psi(r)$ for a RNBH with $Q = 0.09$, $r_H = 0.15$,
  and $\mu = 1 = \lambda$, $\nu = 9/32$ taking different values for $q$.}
\label{fig:Radial1}
\end{center}
\end{figure}

\begin{figure}[h]
\begin{center}
\includegraphics[width=0.5\textwidth]{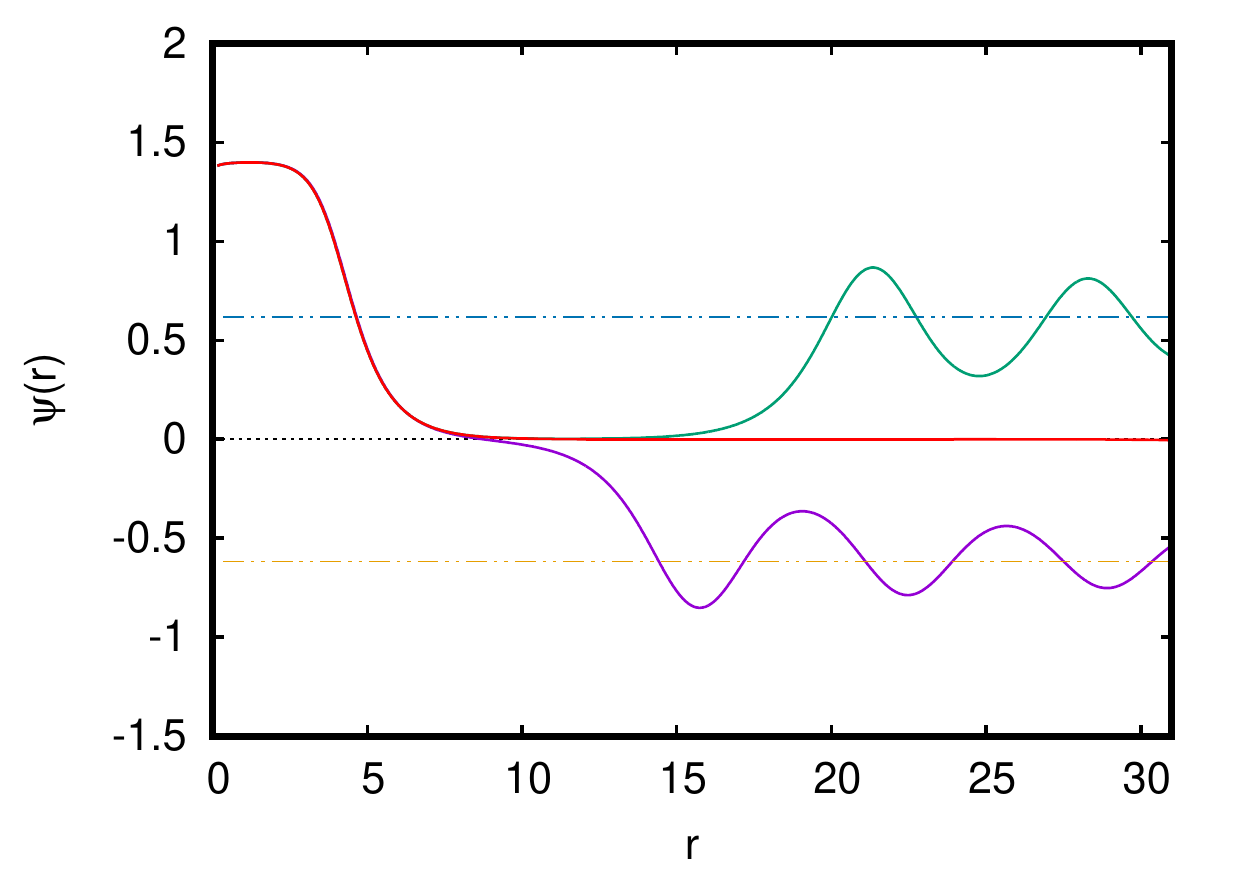}
\caption{Solutions $\psi(r)$ (solid lines) with $q = 1$ for three different values of $\psi_H$
    within a RNBH with $Q = 0.09$, $r_H = 0.15$, $\mu = 1 = \lambda$ and $\nu = 9/32$.
    The three solutions correspond to $\psi_H$: 1.3820 (purple line), 1.381958 (green line) and 1.381959590 (red line).
    The horizontal dotted lines represent the two values $\psi_{\rm max} \approx \pm 0.6175$
    associated with the local maxima of the effective potential $U_{\rm eff}$ depicted in 
    Fig.\ref{fig:Potentialeff} for $q=1$. The solution corresponding to an asymptotically vanishing field, which
is the relevant for the current analysis, is marked in red color.}
\label{fig:Radial2}
\end{center}
\end{figure}

Figure~\ref{fig:Potentialeff} depicts the effective potential (\ref{Ueff}) associated with different values for
$q$ taking $\mu = \lambda = 1$, $\nu = 9/32$, $r_H=0.15$ and $Q=0.09$. 
The shooting method aims at the local minimum of this effective potential
located at $\psi=0$, where $U_{\rm eff}$ vanishes, starting from trial values $\psi_H$ that depend on the value for $q$.
In general a numerical exploration shows that this trial values (assuming only positive ones for concreteness) are such
that $0<\psi_H < \psi_{\rm min}^{+}$ where $\psi_{\rm min}^{+}$ is associated with one of the two global minima of $U_{\rm eff}$
given by $\psi_{\rm min}^{+}$, with
$\psi_{\rm min}^{\pm}=\pm \sqrt{\frac{\lambda}{3\nu}\left[1 +\sqrt{1-\frac{3 \mu_{\rm eff,\infty}^2 \nu}{\lambda^2}}\right]}$.
As we stressed above, the local minimum of $U_{\rm eff}$ is at $\psi=0$ and corresponds to the asymptotic value of the field $\psi(r)$.
Therefore for some $q$ the field $\psi(r)$ must climb one of the local maximum of $U_{\rm eff}$ before reaching the local minimum.
A bad shooting can make the field to oscillate around any of the two $\psi_{\rm max}$ associated with the local maxima of $U_{\rm eff}$
  [cf. Fig.~\ref{fig:Radial2} below] or can make the field to go to $\pm \infty$. As remarked in \cite{Brihaye}, giving $Q$, $\mu$ and $r_H$, the charge $q$ is limited
from above by the condition $\mu_{\rm eff,\infty}^2 \geq 0$, which corresponds to $|q| \leq \mu r_H/|Q|$. On the other hand,
the zeros of $U_{\rm eff}$ are given by $\psi=0$ and $\psi=
\pm \sqrt{\frac{\lambda}{2\nu}\left[1 \pm \sqrt{1-\frac{4 \mu_{\rm eff,\infty}^2 \nu}{\lambda^2}}\right]}$. So when
$\mu_{\rm eff,\infty}^2 = \frac{\lambda^2}{4\nu}$, we see that the charge $q$ is limited from below $ |q|_{\rm min} \lesssim q$ where $|q|_{\rm min}\approx \frac{r_H\mu}{|Q|}\sqrt{1-\frac{\lambda^2}{4\nu \mu^2}}$, assuming $\nu>0$.
All this analysis is qualitative, but gives a fair description of the actual numerical study. In particular,
the lower bound $|q|_{\rm min}$ is approximate, since in this analysis we are neglecting the contribution of the
metric function $N(r)$ in $U_{\rm eff}$ and taking it as if $N(r)=1$. Moreover,
for this particular value of $\mu_{\rm eff,\infty}^2$ the potential $U_{\rm eff}$ ``degenerate'' in that
the two global minima $\psi_{\rm min}^{\pm}$ become also two of its three zeros
at $\psi= \pm \sqrt{\frac{\lambda}{2\nu}}$. In this degenerate situation $\psi(r)= \psi_{\rm min}^{+}$ is an 
approximate solution. So when $|q|\rightarrow |q|_{\rm min}$
the actual positive value $\psi_H\rightarrow \psi_{\rm min}^{+}$, and since in this limit situation $\psi_{\rm min}^{+}$ is
an approximate solution for the field, then the field $\psi(r)$ remains very close to the constant value $\psi_{\rm min}^{+}$
for a relatively large values $r$ and then interpolates to the asymptotic value $\psi=0$ associated with
$\psi(r\rightarrow \infty)$, and the actual numerical solution resembles a step function, as we can appreciate from
Fig.~~\ref{fig:Radial1}. This figure depicts some examples of Q-clouds solutions $\psi(r)$ for different values of $q$
for a RNBH with charge $Q = 0.09$, horizon $r_H = 0.15$, and a scalar field with mass $\mu = 1$.
As $q$ approaches its minimum value, we see that the Q-cloud solution start looking like a step function.
Our results are in agreement with those obtained in \cite{Brihaye}.

\begin{figure}[h]
\begin{center}
\includegraphics[width=0.5\textwidth]{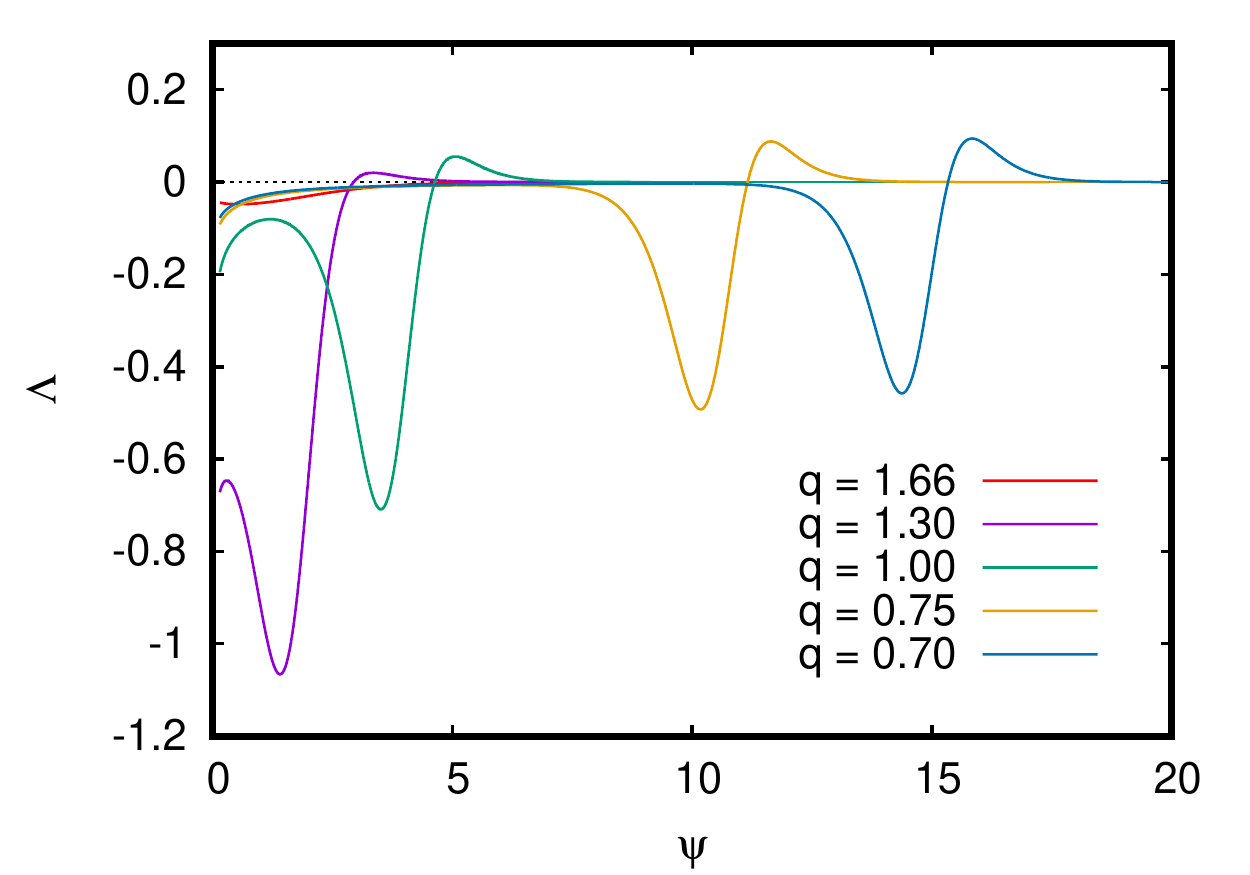}
\caption{The quantity $\Lambda(\psi)$ associated with the solutions of Fig.~\ref{fig:Radial1}. Notice that this quantity
  can be negative and contributes non-trivially to the integral (\ref{Integral4}).}
\label{fig:GammaSubextremal}
\end{center}
\end{figure}

Figure~\ref{fig:GammaSubextremal} shows the quantity $\Lambda(\psi)$
that appears in Eq.~(\ref{Integral4}), when using the solutions $\psi(r)$ that are plotted in Fig.~\ref{fig:Radial1}.
From Fig.~\ref{fig:GammaSubextremal} we appreciate that $\Lambda(\psi)$ has indeed negative contributions to
the integrand of the integral (\ref{Integral4}). The fact that the integrand has negative and positive contributions
allow us to understand why this integral vanishes when $\psi (r)$ is not necessarily the trivial solution
$\psi(r)\equiv 0$, in contrast with the scenario of Sec.~\ref{sec:theorem} where the self-interaction terms are absent leading
to an integrand which is never negative and therefore implying that $\psi(r)\equiv 0$ is the only possible well behaved solution.

Figure~\ref{fig:Radial2} shows three numerical solutions for $\psi(r)$ with $q = 1$ associated with three different (albeit
very similar) values $\psi_H$. The two oscillating solutions correspond to the two values $\psi_H$ that undershoot and
overshoot the desired asymptotic value $\psi=0$, and which asymptotically oscillate around $\psi_{\rm max}$
associated with the local maxima of $U_{\rm eff}$, given by $\psi_{\rm max} \approx \pm 0.6175$
\footnote{If the background were not fixed, 
  the equivalent of those two solutions would led to an spacetime that is not
  asymptotically flat but perhaps asymptotically de Sitter (e.g. if the oscillations falls-off sufficiently fast)
  with an effective cosmological constant given by $U_{\rm eff}^{\rm max}= U_{\rm eff}(\psi_{\rm max})$.}.
These two values are represented by the horizontal dotted lines. The non-oscillating solution 
corresponds to the optimal shooting value $\psi_H$ leading to an asymptotically vanishing solution.

\begin{figure}[h]
\begin{center}
\includegraphics[width=0.5\textwidth]{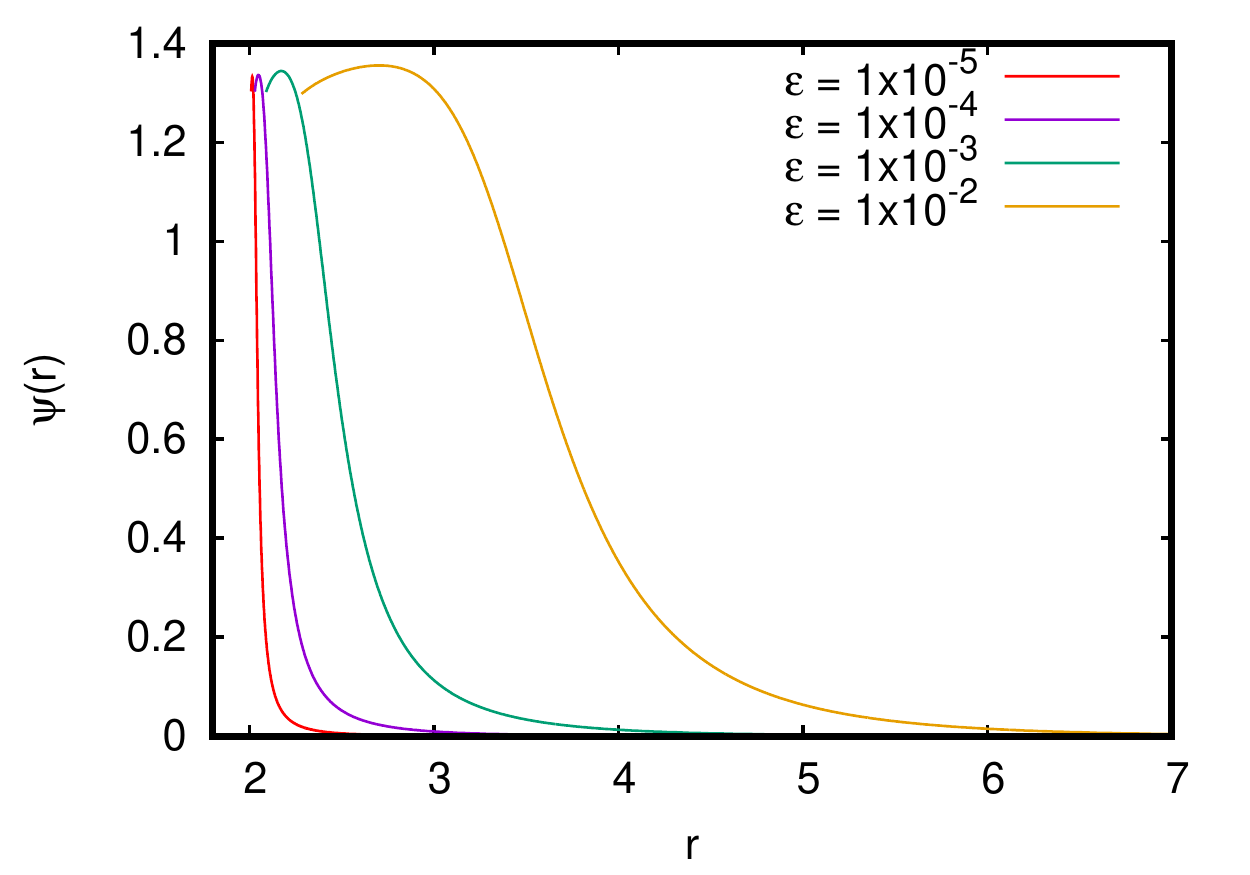}
\caption{Q-cloud solutions $\psi(r)$ for near extremal RNBH $Q \approx M$, taking $M = 2$, $\mu = 1 = \lambda$, $\nu = 9/32$ and $q = 0.8$. The charge of the black hole is taken to be $Q = (1 - \varepsilon)M$ with $\varepsilon = 10^{-2}$, $10^{-3}$, $10^{-4}$ and $10^{-5}$.}
\label{fig:RQuasiExt}
\end{center}
\end{figure}

\begin{figure}[h]
\begin{center}
\includegraphics[width=0.5\textwidth]{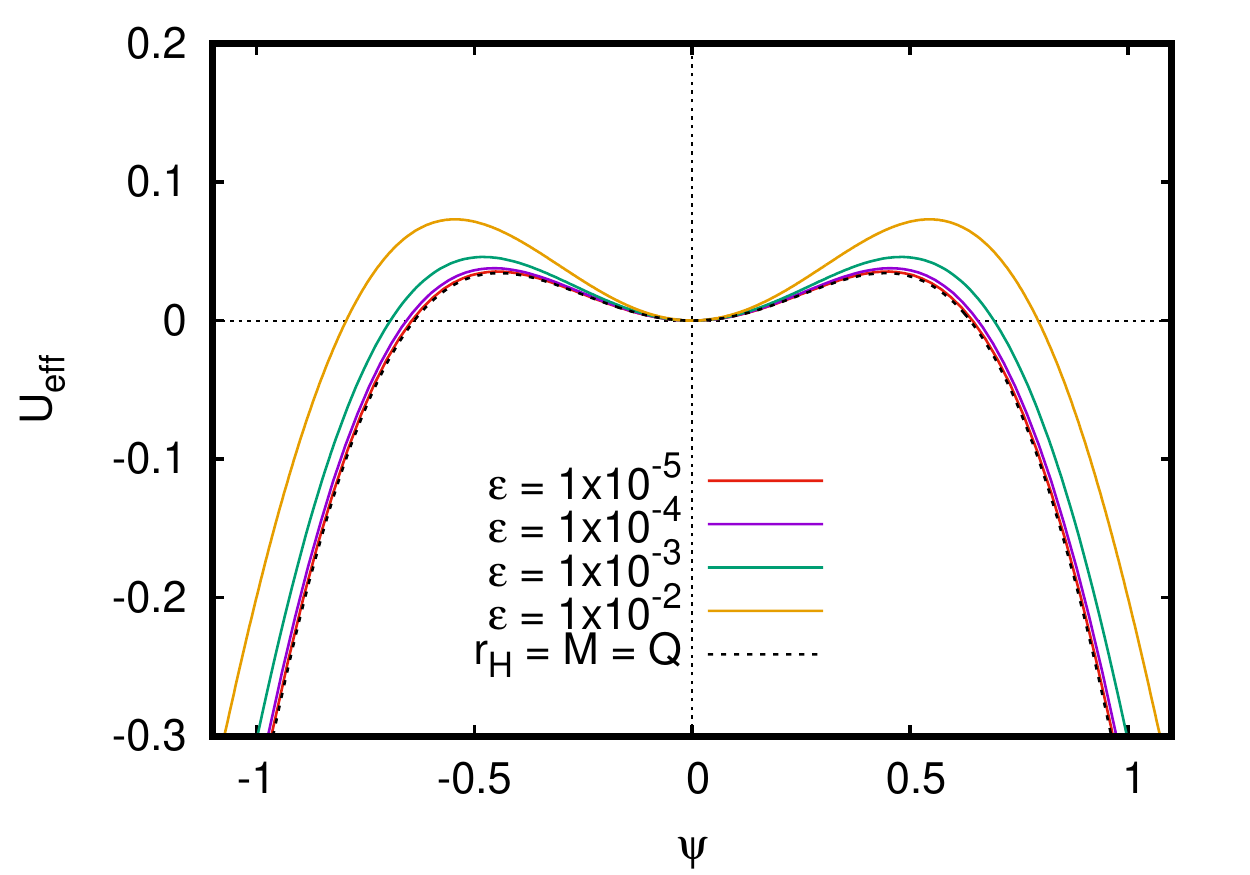}
\caption{The effective potential $U_{\rm eff}$ (\ref{Ueff}) associated with the radial solutions for the near extremal RNBH (Fig.~\ref{fig:RQuasiExt}) taking $\lambda = 1$, $\nu = 9/32$, $\mu = 1$ and $q = 0.8$. The charge of the black hole is taken to be $Q = (1 - \varepsilon)M$ with $\varepsilon = 10^{-2}$, $10^{-3}$, $10^{-4}$ and $10^{-5}$.
  For reference, the exact extremal case $r_H = M = Q $ is depicted by the black dashed line.}
\label{fig:PotentialeffQuasiExt}
\end{center}
\end{figure}

\begin{figure}[h]
\begin{center}
\includegraphics[width=0.5\textwidth]{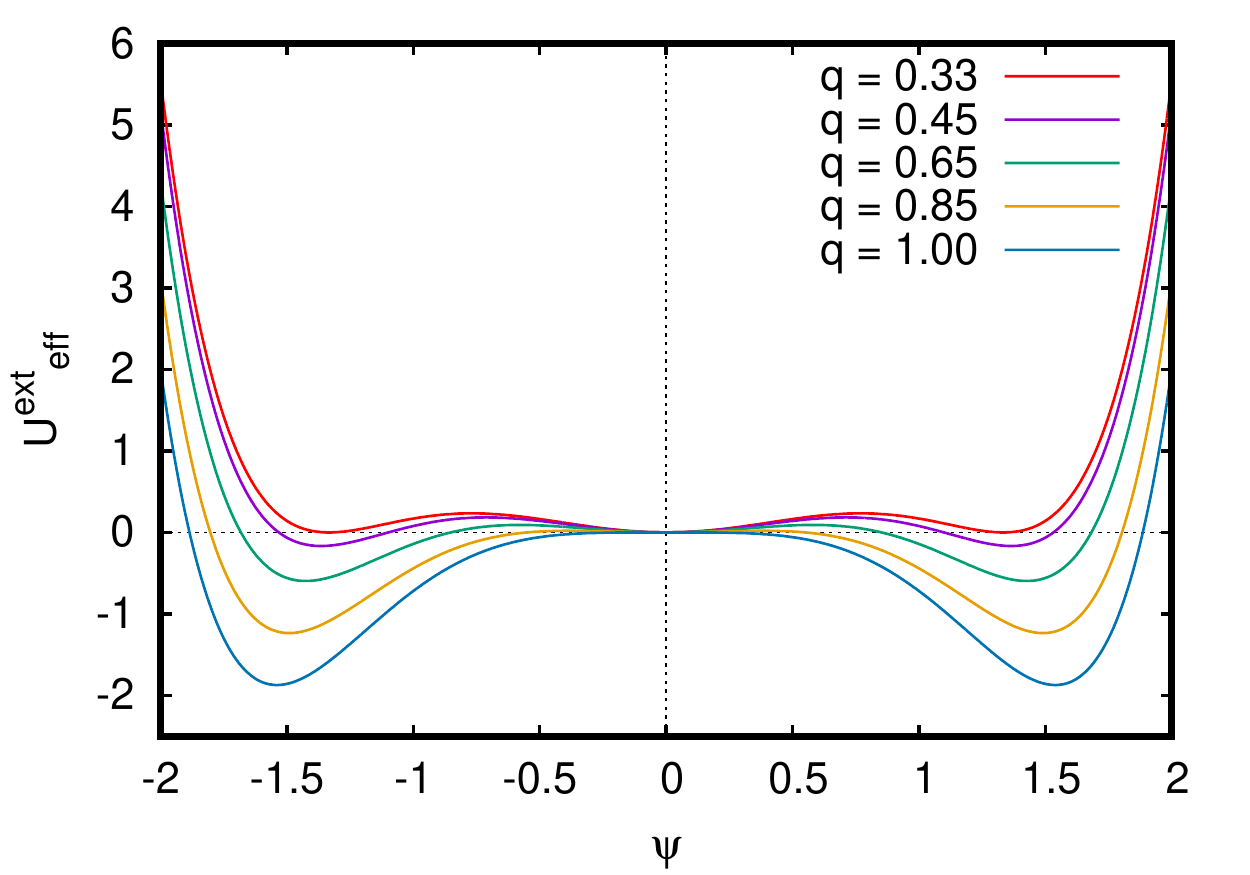}
\caption{The effective potential $U^{\rm ext}_{\rm eff}$ (\ref{Ueff}) taking $\mu = 1 = \lambda$, $\nu = 9/32$,
  and $Q = r_H = 0.15$, for different values of $q$.}\label{fig:PotentialeffExt}
\end{center}
\end{figure}

\begin{figure}[h]
\begin{center}
\includegraphics[width=0.5\textwidth]{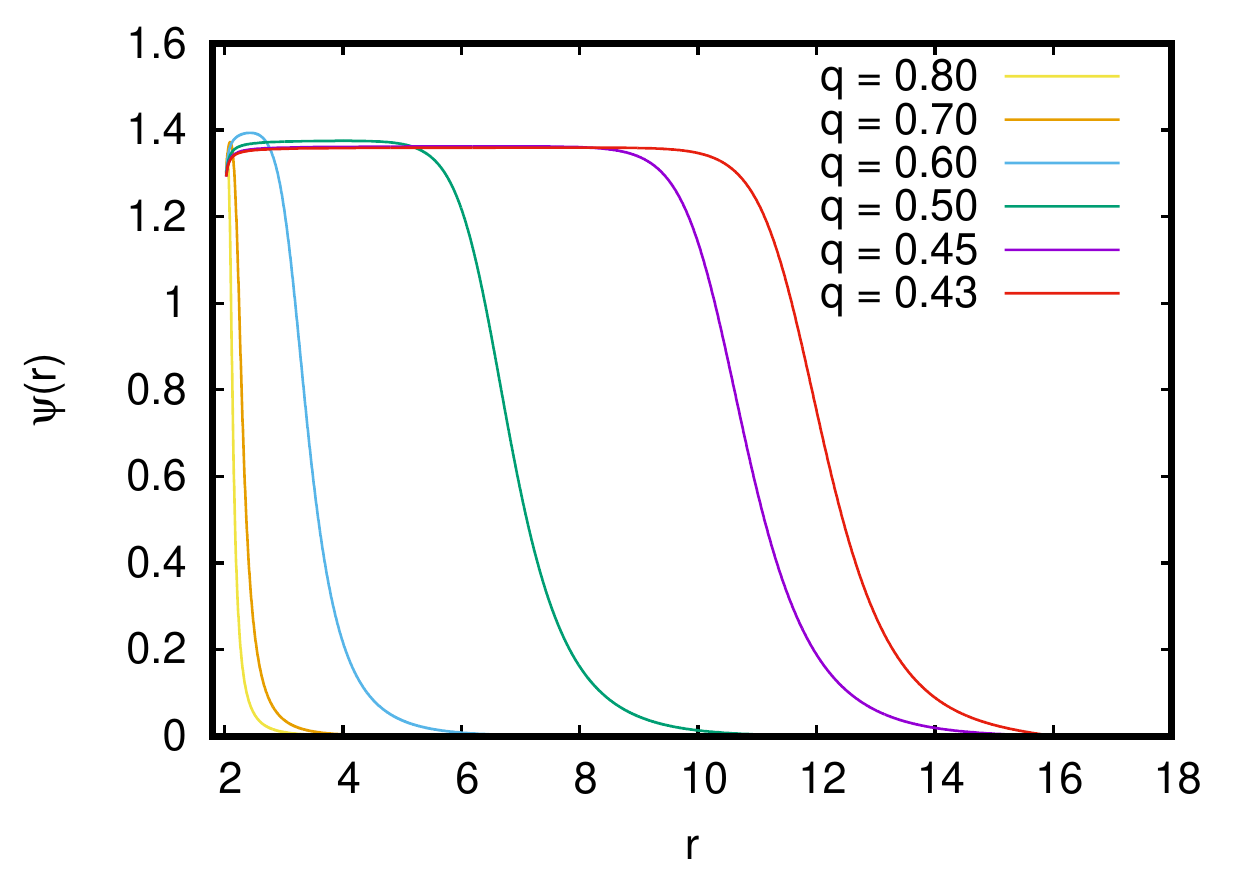}
\caption{Q-cloud solutions $\psi(r)$ for near extremal RNBH
  $Q \approx M$, taking $M = 2$, $\mu = 1 = \lambda$, $\nu = 9/32$.
  The charge of the black hole is taken to be $Q = (1 - \varepsilon)M$ with $\varepsilon = 10^{-4}$ and
  for values $q$ approaching $1/3$.}
\label{fig:RQuasiExt2}
\end{center}
\end{figure}

We have also obtained Q-cloud solutions in
the near extremal RNBH scenario $Q \approx M$ which are consistent with those reported in \cite{Hong2020}. In this
  scenario the charge $q$ is bounded as follows $q_{\rm min} \lesssim q \lesssim \mu$,
  where $q_{\rm min}\approx 1/3$, and this value is obtained from $|q_{\rm min}|$ in the near extremal
  limit and when $\nu$ saturates the bound $\lambda^2/4\mu$ required for the scalar-field potential (\ref{PotentialBH})
  to have a true vacuum at $\Psi=0$. 
  Figure~\ref{fig:RQuasiExt} shows four solutions of this kind as
  $Q$ approaches $M$ and Fig.~\ref{fig:PotentialeffQuasiExt} shows the effective potential associated with these solutions.
  From the regularity conditions (\ref{RegCond1}) and (\ref{RegCond2})
  we appreciate that as $Q \rightarrow M$ the derivatives diverge at the horizon, a feature that can be appreciated also in
  Fig.~\ref{fig:RQuasiExt}. Due to this divergent behavior at the horizon,
  the exact extremal case $(Q = M)$ requires a separate analysis that demands a different numerical technique
  \cite{Garciaetal}. Moreover, this analysis is also necessary to prove that if a non-trivial
  physically meaningful solution exists for the field $\psi(r)$ in the exact extremal scenario, then the kinetic
  term in the integral (\ref{Integral4}), namely $g^{rr} (\psi')^2$, 
  remains well behaved, notably at the horizon, despite a divergent $\psi'_H$. At this respect it is also
  interesting to remark that non-trivial Q-cloud solutions in the extremal scenario with bounded derivatives at the horizon
  are absent \cite{Garciaetal}, and the only ones allowed that have bounded derivatives are the trivial ones
  corresponding to the zeros of $\Lambda_{\rm ext}(\psi)$ (cf. Fig.~\ref{fig:Extremal})
  or equivalently, to the extrema (minima and maxima) of $U_{\rm eff}^{\rm ext}$
  depicted by Fig.~\ref{fig:PotentialeffExt}. Thus, in the near extremal
  case when $q$ approaches $1/3$ something similar happens to the subextremal solutions when
  $q\approx q_{\rm min}$. Namely, the solutions have a ``step function'' shape, where $\psi(r)$ remains
  near the global minima of $U_{\rm eff}$ for larger values of $r$ as
  $q \rightarrow 1/3$, and then interpolates to the local minimum of $U_{\rm eff}$ associated
  with the asymptotic value $\psi\rightarrow 0$ passing through a local maximum. This behavior is
  depicted by Fig.~\ref{fig:RQuasiExt2}.
 

\begin{figure}[h]
\begin{center}
\includegraphics[width=0.5\textwidth]{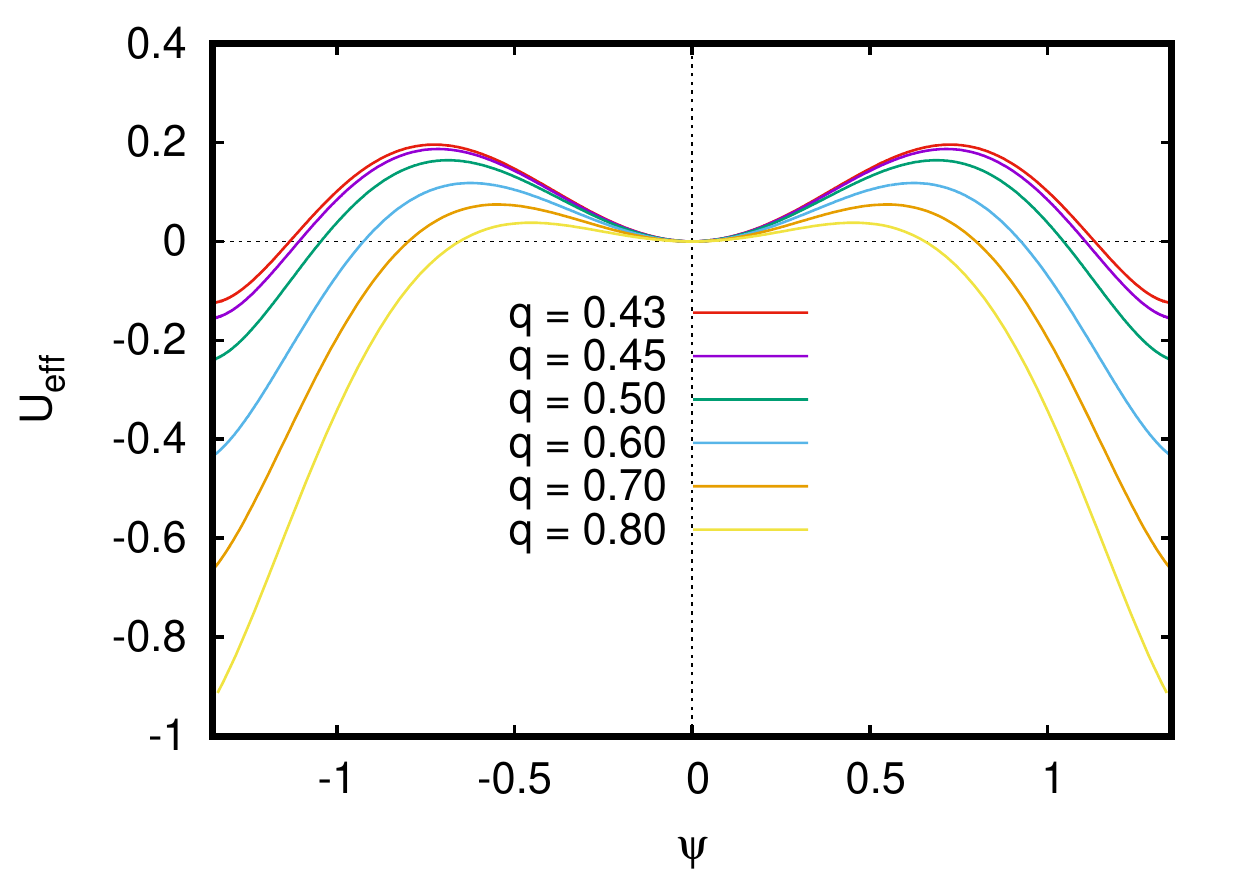}
\caption{The effective potential $U_{\rm eff}$ (\ref{Ueff}) associated with the radial solutions for the near extremal RNBH (Fig.~\ref{fig:RQuasiExt2}) taking $\lambda = 1$, $\nu = 9/32$ and $\mu^{2} = 1$ and
    for different values of $q$ that approach the positive value $q_{\rm min}^{\rm ext}\approx 1/3$
    with $Q = (1 - \varepsilon)M$ and $\varepsilon = 10^{-4}$.}
\label{fig:PotentialeffQuasiExt3}
\end{center}
\end{figure}

\begin{figure}[h]
\begin{center}
\includegraphics[width=0.5\textwidth]{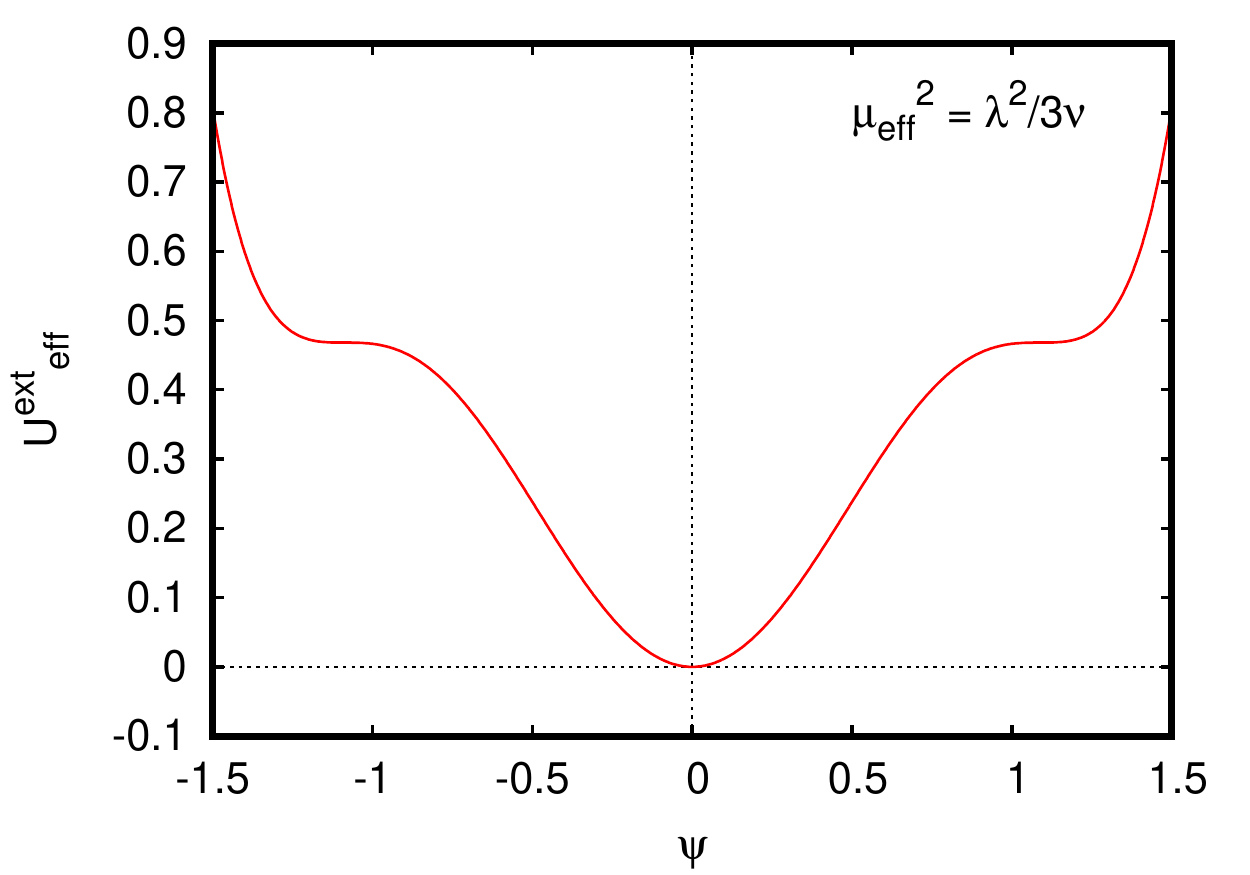}
\caption{The effective potential $U^{\rm ext}_{\rm eff}$ (\ref{Ueff}) for an extremal RNBH ($r_H=|Q|=M$)
  taking $\mu = 1 = \lambda$, $\nu = 9/32$,
  and for $\mu_{\rm eff,ext,\infty}^2= \lambda^2/3\nu= 32/27$.}\label{fig:PotentialeffExt3}
\end{center}
\end{figure}

As mentioned before, in the exact extremal scenario $\mu^2_{\rm eff}(r)$
  given by (\ref{mueffr}) reduces to $\mu^2_{\rm eff}(r)= \mu^2_{\rm eff,\infty,ext}=\mu^2- q^2$ which is not position
  dependent anymore, and then the extrema (minima and maxima) of $U_{\rm eff}^{\rm ext}$ become exact but trivial Q-cloud
  solutions. Moreover, when $\mu^2_{\rm eff, ext,\infty} = \frac{\lambda^2}{3\nu}$,
  the effective potential (\ref{Ueff}) with $\mu^2_{\rm eff} = \mu^2_{\rm eff, ext,\infty}$, has only one global minima
  at $\psi=0$ (see Fig.~\ref{fig:PotentialeffExt3}), which is the only possible solution since in this case
  the function $\Lambda_{\rm ext}(\psi)$ is positive semidefinite and therefore the integral (\ref{Integral4}) only holds
  if $\psi(r)\equiv 0$. In this particular extremal scenario there is not even a
  real valued $q$ that satisfies the condition $\mu^2_{\rm eff, ext,\infty} = \mu^2- q^2=\frac{\lambda^2}{3\nu}$ unless
  $\nu \geq \lambda^2/3\mu^2$.
  
\section{Conclusion}
\label{sec:conclusion}

A massive, charged and complex-valued scalar field coupled to a Reissner-Nordstrom black hole has been studied for the subextremal and extremal scenarios, by imposing regularity conditions on the field, notably, in the radial part.
When the scalar-field potential has no self-interactions terms it
is not possible to find non-trivial {\it superregular} numerical solutions or 
scalar clouds (i.e. solutions where the field and its radial derivatives are bounded in the domain of outer communication
of the BH and at the BH horizon as well).
The extremal scenario with an {\it extremal} scalar field $(|q| = \mu)$, dubbed {\it double extremal}, was previously studied by
Degollado $\&$ Herdeiro \cite{DegolladoHerdeiro2013} reporting non-trivial solutions, but it is unclear to what extent
those solutions are regular at the horizon, given that
using an integral method we have proven a theorem establishing that such non-trivial configurations cannot
exist even if one allows a certain singular behavior on the
radial derivative of the field at the extremal horizon while keeping the {\it kinetic} scalars associated with the
boson field bounded there. Therefore, this conclusion casts doubts about the physical significance of the solutions found in
\cite{DegolladoHerdeiro2013}. On the other hand, by implementing the same integral method to the case where the
scalar-field potential has self-interaction terms, it is not possible to prove a similar theorem, which in turn provides a
heuristic justification and understanding for the existence of regular and spherically symmetric
cloud solutions within this variant of the theory
(termed Q-clouds) which have been reported recently by several authors \cite{Hong2020,Herdeiro2020,Brihaye} and
that we have reproduced here in the background of a subextremal (including a near extremal) RNBH.
Finally, given that the radial derivative of the field may diverge at the BH horizon in the
extremal scenarios (cf. \cite{Hod2012,Hod2015} for
non-charged and charged clouds in the backgrounds of an extremal Kerr and an extremal Kerr-Newman black holes, respectively) while keeping the solutions for the clouds physically meaningful~\cite{Garciaetal}, a more detailed study is in order for the
numerical analysis of Q-clouds in the presence of an exact extremal RN black hole and not only in the near extremal
limit~\cite{Garciaetal}.
\section*{Acknowledgments}
This work was supported partially by DGAPA--UNAM grant IN111719 and
CONACYT (FORDECYT-PRONACES) grant 140630. G.G. acknowledges CONACYT scholarship 291036.    
We are indebted to P. Grandcl\'ement, and E. Gourgoulhon for fruitful discussions and valuable suggestions.


\appendix
\section{Non-existence of bound states for a free scalar field around Reissner-Nordstrom BH}
\label{aped.A}

\bigskip

It is instructive to recover the no-hair theorem presented in Sec.~\ref{sec:theorem} in a much more simplified
fashion. We begin by considering Eq.(\ref{radialE}) for the radial function $R = R(r)$:
\begin{equation}\label{eq.A1}
\left(\Delta R'\right)' = \left[\left(K_l + \mu^2r^2\right) - \frac{\mathcal{H}^2}{\Delta}\right]R ,  
\end{equation}
multiplying both sides by $R$ and integrating both sides from $r_H$ to infinity, we obtain
\begin{equation}\label{eq.A2}
\int_{r_H}^{\infty}R\left(\Delta R'\right)'dr = \int_{r_H}^{\infty}\left[K_l + \mu^2r^2 - \frac{\mathcal{H}^2}{\Delta}\right]R^2dr\,.
\end{equation}
Integrating by parts the left-hand side of previous equation reads
\begin{equation}\label{eq.A3}
\int_{r_H}^{\infty}R\left(\Delta R'\right)'dr = \left. R\Delta R'\right\vert_{r_H}^{\infty} - \int_{r_H}^{\infty}\left(\Delta R'\right)R'dr\,.
\end{equation}
Assuming the following conditions at the horizon and asymptotically [cf. Eq.~(\ref{Rasym})]
\begin{eqnarray}\label{eq.A4}
&& R(r_H), R'(r_H) < \infty \quad \text{(finite values at $r_H$)}, \\
&& R(r \rightarrow \infty), R'(r \rightarrow \infty) \rightarrow 0 
\,,
\end{eqnarray}
and given that $\Delta$ vanishes at the horizon,
\[ \Delta_H \equiv r_H^2 - 2Mr_H + Q^2  = 0,\]
we conclude
\begin{equation}\label{eq.A5}
\left. R \Delta R'\right\vert_{r_H}^{\infty} = 0.
\end{equation}
Therefore Eq.~(\ref{eq.A2}) reduces to
\begin{equation}\label{eq.A6}
\int_{r_H}^{\infty}\Big[ \Delta R'^2 + \alpha(r)R^2\Big] dr = 0,
\end{equation}
where we have defined
\begin{equation}\label{eq.A7}
\alpha(r) \equiv K_l + \mu^2r^2 - \frac{\mathcal{H}^2}{\Delta}.
\end{equation}
We observe that the first term $\Delta R'^2$ is positive semi-definite, however,
the second term $\alpha(r)R^2$ does not have an apparent
definite sign. Nevertheless, below we prove that $\alpha(r)$ is not negative.
Since $\mathcal{H}$ is given by Eq.(\ref{H}) and
\begin{equation}
  \Delta = \left(r - r_H\right)\left(r - r_{-}\right) \,,
  \end{equation}
then
\begin{equation}\label{eq.A8}
\alpha(r) = K_l + \mu^2r^2 - \frac{q^2Q^2}{r_H^2}\frac{\left(r - r_H\right)r^2}{\left(r - r_-\right)}.
\end{equation}
Using
\begin{equation}
\frac{r - r_H}{r - r_-} = 1 + \frac{r_- - r_H}{r - r_-},
\end{equation}
it is possible to rewrite equation (\ref{eq.A7}) as
\begin{equation}
\alpha(r) = K_l + \left(\mu^2 - \omega^2\right)r^2 + \omega^2r^2\frac{\left(r_H - r_-\right)}{\left(r - r_-\right)}\,.
\end{equation}
We now appreciate that the function $\alpha(r)$ is positive
semidefinite for $r \geqslant r_H$ and $\mu^2 \geq \omega^2$, and $r_H\geq r_{-}$, which
includes the extremal scenario. In the subextremal case $r_H > r_{-}$, the function $\alpha(r)$ is strictly positive.
Therefore the integrand that appears in the integral (\ref{eq.A6}) is non-negative,
and for this integral to vanish it is necessary that each term in the integrand vanishes identically for all $r\geq r_H$.
In particular, if $\mu^2> \omega^2$,
\begin{equation}
R'(r) \equiv 0 \quad \text{and} \quad R(r) \equiv 0,
\end{equation}
which leads to the trivial solution $\Psi \equiv 0$. We conclude that there
are no \textit{bound states} when the scalar field is coupled to a subextremal Reissner-Nordstrom BH with
$\mu^2> \omega^2$.
In the extremal case $(r_H = M = Q)$ the function $\alpha(r)$ reduces to
\begin{equation}
\alpha(r) = K_l + \left(\mu^2 - \omega^2\right)r^2,
\end{equation}
where $\omega^2 = q^2$ in this case. Again, if $\mu^2 > \omega^2$, then $\alpha(r)$ is strictly positive for all
$r\geq r_H$, and the same conclusion follows about the absence of non-trivial \textit{scalar clouds}.

Finally, when focused on the scenario studied by Degollado and Herdeiro \cite{DegolladoHerdeiro2013}
about charged scalar clouds in the extremal RNBH with an {\it extremal} test field with $\mu = |q|$ then 
\begin{equation}
\alpha(r) = K_l = l(l + 1) \geqslant 0 \,. 
\end{equation}
Thus $\alpha(r)$ is nonzero except for $l = 0$ (i.e. spherically symmetric clouds), and then
the integral (\ref{eq.A6}) vanishes for all $l\geq 0$ only if the radial function satisfies for all $r\geq r_H=M$:
\begin{equation}
R'(r) \equiv 0 \quad \text{and} \quad R(r) \equiv 0, \quad \text{for} \quad l\neq 0,
\end{equation}
and 
\begin{equation}
R'(r) \equiv 0 \quad \text{and} \quad R(r) \equiv {\rm const.}, \quad \text{for} \quad l = 0\,.
\end{equation}
Since we demand that $R(r\rightarrow \infty)\rightarrow 0$, then $R(r) \equiv 0$ also for $l=0$. 
Thus, even in the particular extremal scenario of Ref.~\cite{DegolladoHerdeiro2013},
our analysis shows that non-trivial clouds are not possible.

The final conclusion is that non-trivial regular bound states for a free scalar field
(massive and charged) in the background of a Reissner-Nordstrom black hole, extremal or subextremal, are absent.


\end{document}